\documentclass[12pt]{article}
\usepackage{amsmath, amssymb, amsthm}
\usepackage{graphicx}
\usepackage{enumerate}
\usepackage[numbers]{natbib}
\usepackage{url} % not crucial - just used below for the URL 
\usepackage{subcaption}
\usepackage{algorithm}
\usepackage{algpseudocode}
\usepackage{algcompatible}
\usepackage{hyperref}
\newtheorem{theorem}{Theorem}
\newtheorem{corollary}{Corollary}[theorem]

\newtheorem{proposition}[theorem]{Proposition}
\newcommand{\B}[1]{\boldsymbol{#1}}

\usepackage{doi}
\usepackage{setspace}

\DeclareMathOperator*{\argmax}{argmax}

\newcommand{\btheta}{\boldsymbol{\theta}}

\newcommand{\bpi}{\boldsymbol{\pi}}

\newcommand{\bSigma}{\boldsymbol{\Sigma}}
\newcommand{\bPsi}{\boldsymbol{\Psi}}

\newcommand{\bmu}{\boldsymbol{\mu}}

\newcommand{\bTheta}{\boldsymbol{\Theta}}

\newcommand{\bM}{\boldsymbol{M}}

\newcommand{\bA}{\boldsymbol{A}}
\newcommand{\bS}{\boldsymbol{S}}

\newcommand{\bL}{\boldsymbol{L}}

\newcommand{\bs}{\boldsymbol{s}}

\newcommand{\bx}{\boldsymbol{x}}
\newcommand{\bX}{\boldsymbol{X}}
\newcommand{\by}{\boldsymbol{y}}
\newcommand{\bY}{\boldsymbol{Y}}
\newcommand{\bZ}{\boldsymbol{Z}}

\newcommand{\mL}{\mathcal L}
\newcommand{\mM}{\mathcal M}

\newcommand{\mS}{\mathcal S}

\newcommand{\mA}{\mathcal A}

\newcommand{\mW}{\mathcal W}
\newcommand{\mN}{\mathcal N}

\newcommand{\blind}{1}

\begin{document}

\def\spacingset#1{\renewcommand{\baselinestretch}%
{#1}\small\normalsize} \spacingset{1}

%%%%%%%%%%%%%%%%%%%%%%%%%%%%%%%%%%%%%%%%%%%%%%%%%%%%%%%%%%%%%%%%%%%%%%%%%%%%%%

\if1\blind
{
  \title{\bf Statistical few-shot learning for large-scale classification via parameter pooling}
  \author{Andrew Simpson\hspace{.2cm}\\
    Mathematics and Statistics Department, South Dakota State University\\
    and \\
   Semhar Michael \thanks{
   	The authors gratefully acknowledge \textit{the National Science Foundation and National Geospatial-Intelligence Agency} for their support 
   	%please remember to list all relevant funding sources in the unblinded version
   } \\
    Mathematics and Statistics Department, South Dakota State University}
  \maketitle
} \fi

\if0\blind
{
  \bigskip
  \bigskip
  \bigskip
  \begin{center}
    {\LARGE\bf Statistical few-shot learning for large-scale classification via parameter pooling}
\end{center}
  \medskip
} \fi

\bigskip
\begin{abstract}
%The text of your abstract. 200 or fewer words.
In large-scale few-shot learning for classification problems, often there are a large number of classes and few high-dimensional observations per class. Previous model-based methods, such as Fisher's linear discriminant analysis (LDA), require the strong assumptions of a shared covariance matrix between all classes. Quadratic discriminant analysis will often lead to singular or unstable covariance matrix estimates. Both of these methods can lead to lower-than-desired classification performance. We introduce a novel, model-based clustering method that can relax the shared covariance assumptions of LDA by clustering sample covariance matrices, either singular or non-singular. In addition, we study the statistical properties of parameter estimates. This will lead to covariance matrix estimates which are pooled within each cluster of classes. We show, using simulated and real data, that our classification method tends to yield better discrimination compared to other methods.

%In classification problems when working in a high dimensional and large number of classes with few observations per class, linear discriminant analysis (LDA) requires the strong assumptions of a shared covariance matrix between all classes and quadratic discriminant analysis leads to singular or unstable covariance matrix estimates. Both of these can lead to lower than desired classification performance. We introduce a novel, model-based clustering method which can relax the shared covariance assumptions of LDA by clustering sample covariance matrices, either singular or non-singular. This will lead to covariance matrix estimates pooled within each cluster. We show using simulated and real data our method for classification tends to yield better discrimination compared to other methods.

\end{abstract}

\noindent%
\vfill

\newpage
\spacingset{1.9} % DON'T change the spacing!
\section{Introduction}
\label{Introduction}

An increasing body of problems in statistics and pattern recognition show high-dimensional data with few realizations per class, in addition to a large number of classes. This particularly arises in forensic science and statistics problems \cite{dettmanforensic2014, simpson2024modeling} where goal is to provide a decision maker with a summary that a piece of evidence arose from some source which is commonly referred to as a value of evidence \cite{aitken2004statistics,ommenandsaunders21,  ramosreliable2013} which has been widely discussed in the literature \cite{bergerlr2016, biedermannreframing2016, leegwaterperformance2017, martireinterpretation2014, morrisoncomparison2011, ausdemoretwostage2021, nordgaardlikelihood2012}. 

In forensic science, it may often be assumed that the number of sources, or class, tends to infinity. For example, the evidence may be a copper wire from an improvised explosive device, and the possible class that gave rise to that piece of copper wire is the set of spools of copper wire, which is extremely large. In this framework, since there are only a few observations per class in a high-dimensional space, there are often not enough observations to estimate a non-singular and stable covariance estimate under the assumptions of normality. 

Hence, parameter pooling assumptions are made, such as assuming that each class shares the same covariance matrix. In the classification setting, this is analogous to Fisher's linear discriminant analysis (LDA) \cite{rencher2005review}, and assuming each class has a unique covariance matrix is analogous to quadratic discriminant analysis (QDA) \cite{rencher2005review}. The methods proposed in this paper aim to find a middle ground between LDA, which suffers from strong assumptions, and QDA, which often has a degenerate solution when there are few observations per class. 

In this paper, the parameter pooling approach for few-shot learning is considered. This is done by assuming there is a set of $K$ latent covariance matrices that are shared between all classes as opposed to just a single covariance matrix shared between all classes as in LDA or as opposed to QDA where the number of shared covariance matrices is equal to the number of classes. This is done by leveraging the fact that a large number of classes exist. Under the assumptions of normality of each class, this is achieved via model-based clustering of singular and non-singular sample covariance estimates. 

While there is a wide breadth of literature in model-based clustering assuming the observations being clustered live in a Euclidean space \cite{mclachlanfinite2000, bouveyronmodel-based2014, celeuxgaussian1995, fraleymodel-based2002, melnykovfinite2010, melnykovclustering2020}, there are fewer when the observations being clustered are covariance matrices. \citet{hidotexpectationmaximization2010} considered the case of clustering non-singular symmetric matrices but does not address the case of clustering singular symmetric matrices. 

In this paper, clustering singular symmetric matrices and non-singular symmetric matrices at the same time will be addressed. \citet{rodriguez2024improving} took a similar approach with covariance clustering but on a small number of classes and in a non-few-shot setting.  \citet{rayens1991covariance} have attempted to correct for the downsides of LDA and QDA by taking a linear combination of the within-class sample covariance estimate and the pooled covariance estimate but introduce a regularization hyper-parameter that must be picked.By considering the existence of a finite set of covariance matrices shared between all classes, we introduce the latent covariance discriminant analysis (LCDA) when used for classification.. 

The rest of the paper is organized as follows. Section~\autoref{Methods} introduces the respective models and methods for finding parameters when singular matrices exist, when non-singular matrices exist, and when both singular and non-singular matrices exist. Along with this, multiple properties are given just at consistency and deriving the optimal Bayes rule. Section~\ref{Simulated Data} provides an empirical study for the proposed method to study various aspects. Section~\ref{Real Data} applies the proposed method LCDA to a glass and handwriting dataset and compares out-of-sample classification accuracy with LDA and, when possible, with QDA..

\section{Statistical few-shot learning methodology}
\label{Methods}
This section details the few-shot statistical learning model, estimation approaches, and statistical properties. In Section~\ref{sec: EM} we state the preliminary on finite mixture models and the EM algorithm.  In Section~\ref{sec:prelim} we introduce the problem and notations followed by three approaches for few-shot model pooling and their respective algorithms in Section~\ref{sec: wishart}~-~\ref{sec: normal}. Theoretical results are given in Section~\ref{sec: theory}.

\subsection{Finite mixture models and estimation}%Expectation Maximization Algorithm for Finite Mixture Models}
\label{sec: EM}
Consider a set of some random vectors $\bY^n = \{\bY_1, \bY_2, \ldots, \bY_n\}$ and let finite set of $K$ distributions with probability density functions (pdf) $f_1, f_2, \dots, f_K$ parameterized by ${\btheta}_1, {\btheta}_2, \dots, {\btheta}_K$, respectively, describe mixture components. Let $Z_i$ denote the distribution/component which generated $\bY_i$ where $\bY_i | Z_i = k \sim f_k$ which then implies the unconditional pdf of $\bY_i$ is
\begin{math}
    f(\by; {\bPsi}) = \sum_{i=1}^n\pi_k f_k(\by; {\btheta}_k)
\end{math}
where $\pi_k = P(Z_i=k)$ is the prior probability that $\bY_i$ was generated by the $k^{th}$ component and ${\bPsi} = \{{\btheta}_1, {\btheta}_2, \dots, {\btheta}_K\}$. It is said that $\bY_i$ follows a finite mixture model detonated as $\bY_i \sim FMM({\bPsi})$. The maximum likelihood estimate (MLE) for the parameter vector $\bPsi$ is defined as
\begin{math}
    \hat{{\bPsi}} = \argmax_{{\bPsi}}\mL({\bPsi}|\by^n) = \argmax_{{\bPsi}}\ell({\bPsi}|\by^n),
\end{math}
where $\mL({\bPsi}|\by^n)$  and $\ell({\bPsi}|\by^n)$ are the likelihood and the log-likelihood function defined respectively as

\begin{equation}
     \mL({\bPsi}|\by^n) = \prod_{i=1}^n \sum_{k=1}^K \pi_k f_k(\by_i; {\btheta}_k)
\end{equation}

and

\begin{equation}
    \ell({\bPsi}|\by^n) = {\log}\mL({\bPsi}|\by^n) = \sum_{i=1}^n {\log}\left(\sum_{k=1}^K \pi_k f_k(\by_i; {\btheta}_k)\right)
\end{equation}
with $\by^n = \{\by_1, \by_2, \ldots, \by_n\}$ denoting realizations.
For most problems, it can be difficult to find closed-form solutions for parameter estimates $\hat{{\bPsi}}$ that maximize $\ell({\bPsi}|\by^n)$. Therefore, the expectation maximization (EM) algorithm \cite{apdempstermaximum1997} is used. The EM algorithm is an iterative algorithm that maximizes $\ell({\bPsi}|\by^n)$ to achieve the maximum likelihood estimates of $\bPsi$ assuming proper initialization \cite{michaeleffective2016}. To achieve this, the complete-data \text{log}-likelihood is defined as the \text{log}-likelihood given the observations as well as the unobserved random variables, which in this case are ${\bZ} = \{Z_1, Z_2, \dots, Z_n \}$. The complete-data \text{log}-likelihood is then
\begin{math}
    \ell_c({\bPsi}|\by^n, {\bZ}) = \sum_{i=1}^n \sum_{k=1}^K I(Z_i = k) (\log\pi_k + \log f_k(\by_i; {\btheta}_k)),
\end{math}
where $I(.)$ is the indicator function. The EM algorithm has two parts. The E-step finds the conditional expectation of $\ell_c({\bPsi}|\by^n)$ given the observations and the parameter estimates from the previous step. This is commonly referred to as the $\mathcal{Q}-$function given as
\begin{equation*}
    \mathcal{Q}({\bPsi}|\by^n, {\bZ}, {\bPsi}^{(t)}) = E_{\bPsi}(\ell_c({\bPsi}|\by^n, {\bZ}) | \by^n; {\bPsi}^{(t)}) = \sum_{i=1}^n \sum_{k=1}^K \tau_{ik}^{(t)} (\log\pi_k + \log f_k(\by_i; {\btheta}_k)),
\end{equation*}
where $\tau_{ik}^{(t)} = E_{\bPsi}(I(Z_i = k) | \by^n; {\bPsi}^{(t)}) = P(Z_i = k |\by_i; {\bPsi}^{(t)}) = \frac{\pi_k f_k(\by_i; \btheta_k^{(t-1)})}{\sum_{k'=1}^K \pi_k f_{k'}(\by_i; {\btheta_{k'}^{(t-1)}})}$, which is often (t)referred to as the posterior probability of the membership of the $i$th observation to the $k$th component. The M-step consists of maximizing $\mathcal{Q}({\bPsi}|\by^n, {\bZ}, {\bPsi}^{(t)})$ to get new parameter estimates for the next iteration by
\begin{math}
    {\bPsi}^{(t+1)} = \argmax_{{\bPsi}} \mathcal{Q}({\bPsi}|\by^n, {\bZ}, {\bPsi}^{(t)}).
\end{math}
Deriving ${\bPsi}^{(t+1)}$ will depend of the distributional assumptions $f_1, f_2, \dots, f_K$. Given the EM algorithm is iterative, it must be given initial values ${\bPsi}^{(0)}$ at which point the E-step and M-step are iterated until some convergence creation is met, for example, until $\frac{\mL({\bPsi}^{(t)}| \by^n) - \mL({\bPsi}^{(t-1)}| \by^n)}{|\mL({\bPsi}^{(t-1)}| \by^n)|} < \epsilon$ for some small $\epsilon > 0$. After the convergence criteria have been met, the final iteration parameter values are taken as the parameter estimate $\hat{{\bPsi}}$. Once parameter estimates are found, one can estimate the component/distribution which gave rise to some observation by the maximum a posteriori rule (MAP) in which we have $\hat{Z}_i = \argmax_k P(Z_i = k | y_i; \hat{{\bPsi}}) = \argmax_k \tau_{ik}$. This base pseudo-code is outlined in Algorithm~\autoref{alg:Mixture_EM}.

\begin{algorithm}[H]
\caption{EM Algorithm for Finite Mixture Models}
\label{alg:Mixture_EM}
\begin{algorithmic}
\setstretch{1.5}

\STATE \textbf{Input:} Observed data $\by = \{y_i\}_{i=1}^{n}$, number of components $K$, convergence threshold $\epsilon > 0$
\STATE \textbf{Initialize:} Parameter set ${\bPsi}^{(0)} = \{\pi_k^{(0)}, \btheta_k^{(0)}\}_{k=1}^{K}$, set iteration counter $t \gets 0$

\WHILE {$\frac{\mathcal{L}({\bPsi}^{(t)}) - \mathcal{L}({\bPsi}^{(t-1)})} {|\mathcal{L}({\bPsi}^{(t-1)})|} \geq \epsilon$}

    \STATE \textbf{E-Step:} Compute the posterior probabilities (responsibilities):
    
   $ \tau_{ik}^{(t)} \gets \frac{\pi_k^{(t)} f_k(y_i \mid \btheta_k^{(t)})}
    {\sum_{k'=1}^{K} \pi_{k'}^{(t)} f_{k'}(y_i \mid \btheta_{k'}^{(t)})}
   $ 
    \quad for all $i \in \{1, \dots, n\}$ and $k \in \{1, \dots, K\}$.

    \STATE \textbf{M-Step:} Update parameter estimates:
    
   $ {\bPsi}^{(t+1)} \gets \argmax_{{\bPsi}} \mathcal{Q}({\bPsi} \mid \by, {\bPsi}^{(t)})
   $ 

    \STATE Increment iteration counter: $t \gets t+1$
    
\ENDWHILE

\STATE \textbf{Output:} Converged parameter estimates $\hat{\bPsi} \gets {\bPsi}^{(t)}$
\STATE \textbf{return} $\hat{\bPsi}$

\end{algorithmic}
\end{algorithm}

\subsection{The statistical few-shot learning problem setup}
\label{sec:prelim}
Consider a classification problem with $n$ classes and $n_i$ observation within each class. Since our problem setup is in the large-scale few-shot classification problem, we assume that we have a large number of classes with few high dimensional observations per class. Suppose random vectors ${\bX}_{ij} | Z_i = k \sim N_p({\bmu}_i, {\bSigma}_k)$ for $i \in \{1,2, \ldots, n\}$ classes, $j \in \{1,2, \ldots, n_i\}$ observations within class $i$, and $k =1, \ldots, K$ clusters of covariance matrices. Here $\bX_{ij} \in \mathbb{R}^p$ is the $j$th observation from the $i$th class, with random variable $Z_i \sim \text{ Categorical}(\pi_1, \pi_2, \dots, \pi_K)$ indicating the covariance matrix of the $i$th class, $N_p$ denoting the multivariate normal distribution on the space $\mathbb{R}^p$ where ${\bmu}_i \in \mathbb{R}^p$ and ${\bSigma}_k \in \mathbb{R}^{p \times p}$ denoting the mean vector and the covariance matrix, respectively. Henceforth, we will refer to this model as the latent covariance model. The latent covariance model is contrasted with linear discriminant analysis (LDA) that assumes ${\bX}_{ij} \sim N_p({\bmu}_i, {\bSigma})$ % for $i \in \{1,2, \dots, n\}$ and $j \in \{1,2, \dots, n_i\}$
and quadratic discriminant analysis (QDA) which assumes ${\bX}_{ij} \sim N_p({\bmu}_i, {\bSigma}_i)$ for a given observation $\bX_{ij}$. % for $i \in \{1,2, \dots, n\}$ and $j \in \{1,2, \dots, n_i\}$
 The goal of this section is to develop adequate methods for estimating the parameter vector ${\bPsi} = \{\pi_1, \dots, \pi_{K-1}, {\bmu}_1, \dots, {\bmu}_n, {\bSigma}_1, \dots, {\bSigma}_K \}$. % which are the parameters of the latent covariance model. 
 As will be seen, finding parameter estimates for $\{{\bmu}_1, \dots, {\bmu}_n\}$ is trivial. The next step is to find estimates for $\bPsi_{\bSigma} = \{{\bSigma}_1, \dots, {\bSigma}_K \}$ and $\bPsi_{\bpi} = \{\pi_1, \dots, \pi_{K-1} \}$. Recall that covariance matrix for each class is commonly estimated using the sample covariance matrix given by $ \hat{{\bSigma}}_i = \frac{1}{n_i-1}\sum_{j=1}^{n_i}({\bX}_{ij} - \bar{{\bX}}_i)({\bX}_{ij} - \bar{{\bX}}_i)^T$ where $\bar{{\bX}}_i$ is the sample mean vector of the $i$th class. Then we can assume that each $\hat{{\bSigma}}_i$ to come from one of $K$ groups with respect to the latent variable $Z_i$, then parameter estimates for $\bPsi_{\bSigma}$ can be found using the expectation maximization (EM) algorithm \cite{apdempstermaximum1997}. If the $\text{rank}(\hat{{\bSigma}}_i) = p$ then $\hat{{\bSigma}}_i \in \mathcal{PD}(p)$, where $\mathcal{PD}(p)=\{\bM \in \mathbb{R}^{p\times p} : \bx^T \bM \bx > 0 \text{ } \forall \bx \in \mathbb{R}^p \}$ is a space of positive definite matrices. Thus, the random matrices that are being clustered are from $\mathcal{PD}(p)$. Since the EM algorithm is being utilized, a density must be defined for $\hat{{\bSigma}}_i$ on $\mathcal{PD}(p)$. Recall that $\bS_i | Z_i$ follows a Wishart distribution, where the scatter matrix is defined as $\bS_i = (n_i - 1) \hat{{\bSigma}}_i$ under the assumption of normality of $\bX_{ij}|Z_i$. If $\text{ rank}(\hat{\bSigma}_i) = a < p$ then $\hat{\bSigma}_i \in \mathcal{PSD}(p, a)$, where $\mathcal{PSD}(p,a)=\{\bM \in \mathbb{R}^{p\times p}: \bx^T \bM \bx > 0 \text{ } \forall \bx \in \mathbb{R}^p, \text{ rank}(\bM) = a\}$ is a space of positive semidefinite matrices. Although the probability density function of $\bS_i | Z_i$ no longer exists on $\mathcal{PD}(p)$, it was shown that a density function exists on $\mathcal{PSD}(p, a)$ and follows a singular Wishart distribution\cite{uhligsingular1994}. In both cases of full rank ($\text{rank}(\hat{\bSigma}_i) = p \text{ } \forall i \in \{1, \dots, p\}$), or singular ($\text{rank}(\hat{{\bSigma}}_i) = a < p \text{ } \forall i \in \{1, \dots, n\}$) covariance matrices, the expectation-maximization algorithm can be derived to find parameter estimates for $\bPsi_{\bSigma}$. These are derived in Sections~\ref{sec: wishart} and \ref{sec: singular_wishart}, respectively. %, since the density of the observations in question $\hat{{\bSigma}}_i$ all exist on the same space of either $\mathcal{PD}(p)$ or $\mathcal{PSD}(p, a)$. This becomes less clear if $\text{rank}(\hat{{\bSigma}}_i) \in \{1, \dots, p\}$ as different observations live in different spaces. 
 In Section~\ref{sec: normal}, a third approach to estimating $\bPsi_{\bSigma}$ is provided which is able to handle the case where $\text{rank}(\hat{{\bSigma}}_i) \in \{1, \dots, p\}$. In the last case, we also show that the parameter estimates of $\bPsi_{\bSigma}$ are asymptotically inconsistent, and a corrected estimate is provided. Furthermore, the optimal Bayes rule for classification is derived. %First, a general derivation of the EM algorithm is given for the case of finite mixture models in Section~\ref{sec: EM}.
 
\subsection{Wishart-based mixture estimation} \label{sec: wishart}

%Consider the latent covariance model ${\bX}_{ij} | Z_i = k \sim N_p({\bmu}_i, {\bSigma}_k)$ for $i \in \{1,2, \dots, n\}$ and $j \in \{1,2, \dots, n_i\}$
%where ${\bX}_{ij} \in \mathbb{R}^p$ is the $jth$ observation from the $ith$ class, $Z_i \sim Categorical(\pi_1, \pi_2, \dots, \pi_K)$ indicates the covariance matrix of the $ith$ class, ${\bmu}_i \in \mathbb{R}^p$, and ${\bSigma}_k \in \mathbb{R}^{p \times p}$ are the respective parameters. 

Let ${\bS}_i = \sum_{j=1}^n({\bX}_{ij} - \bar{{\bX}}_i)({\bX}_{ij} -  \bar{{\bX}}_i)^T$ be the scatter matrix which is the sample covariance estimate of the $ith$ class multiplied by $n_i-1$. Suppose that $\text{rank}(\bS_i) = p \text{ } \forall i \in \{1,\dots, n\}$ where ${\bS}_i \in \mathcal{PD}(p)$. It then follows that ${\bS}_i | Z_i = k \sim \mathcal{\mW}({\bSigma}_k, n_i-1)$ where $\mathcal{W}$ denotes the Wishart distribution and has the centering matrix ${\bSigma}_k$ with $n_i-1$ degrees of freedom defined on the space $\mathcal{PD}(p)$. The density of ${\bS}_i$ conditional on $Z_i$ is then the density of the Wishart distribution
\begin{equation}\label{wishart}
    f_{\mW}({\bs}; {\bSigma}_k, \nu) = \left(2^{\frac{\nu p}{2}} |{\bSigma}_k|^{\frac{\nu }{2}} \Gamma_p\left(\frac{\nu }{2}\right)\right)^{-1}|{\bs}|^{\frac{1}{2}(\nu -p-1)}e^{-\frac{1}{2}tr({\bSigma}_k^{-1}{\bs})},
\end{equation}
where $\nu$ is the degrees of freedom and $\Gamma_p$ is the multivariate gamma function defined as
\begin{math}
    \Gamma_p(x) = \pi^{{p(p-1)}/{4}}\prod_{l=1}^p \Gamma\left( x - \frac{l-1}{2} \right).
\end{math}
%Given the density is defined on $\mathcal{PD}(p)$, it can quickly be seen that if ${\bA} \in \mathcal{PSD}(p,a)$ then $|{\bA}| = 0$ since some of the eigenvalues of $A$ are zero and thus $f_{\mW}({\bA}; {\bSigma}_k, n_i-1) = 0$.
% It must be noted the Wishart distribution only has a density on the space of full \text{rank} positive definite matrices i.e. on the space of $p\times p$ positive definite matrices ${\bM}$ such that $\text{\text{rank}}({\bM})=p$. Thus we must have that $\text{\text{rank}}({\bS})=p$ which is true if $n \ge p$. If it is the case that $n < p$, then there is no density on the space of full \text{rank} matrices but the density does exist in the space of semidefinite matrices ${\bM}$ such that $\text{\text{rank}}({\bM})=n$ \cite{uhligsingular1994}.
Then the unconditional distribution of ${\bS}_i$ follows the finite mixture model given by
\begin{math}
    f({\bs}; {\bTheta}) = \sum_{i=1}^n\pi_k f_{\mW}({\bs}; {\bSigma}_k, n_i-1),
\end{math} 
where the parameter vector $\bTheta = \{\pi_1, \ldots, \pi_K, \bSigma_1, \ldots, \bSigma_K\}$. Denoting ${\bs}^n = \{{\bs}_1, \dots, {\bs}_n\}$, the likelihood and complete-data log-likelihood functions are
% \begin{equation*}
%     \mL_{W}({\bTheta}|s^n) = \prod_{i=1}^n \sum_{k=1}^K \pi_k f_{\mW}(s_i; {\bSigma}_k, n_i).
% \end{equation*}
\begin{equation} \label{eq:wishart_likelihood}
    \mL_{\mW}({\bTheta}|{\bs}^{n}) = \prod_{i=1}^n \sum_{k=1}^K \pi_k f_{\mW}({\bs}_i; {\bSigma}_k, n_i-1)
\end{equation}
and 
\begin{math}
    \ell_c({\bTheta}|{\bs}^n, Z) = \sum_{i=1}^n \sum_{k=1}^K I(Z_i = k) \log(\pi_k f_{\mW}({\bs}_i; {\bSigma}_k, n_i-1)),
\end{math}
respectively.
Thus,  the results  from the E-step of the base EM-algorithm at $t$th iteration can be modified as
\begin{align*}
  \tau_{ik}^{(t)} = P(Z_i = k | {\bs}_i; {\bTheta}^{(t)})
             % &= \frac{ \pi_k^{(t)} f({\bs}_i | Z_i = k; {\bTheta}^{(t)})}{\sum_{k'=1}^K \pi_{k'}^{(t)} f({\bs}_i | Z_i = k'; {\bTheta}^{(t)})} \\
              %&= \frac{ \pi_k^{(t)} f_{\mW}({\bs}_i; {\bSigma}_k, \nu )}{\sum_{k'=1}^K \pi_{k'}^{(t)} f_{\mW}({\bs}_i; {\bSigma}_{k'}, \nu )} \\
              &= \frac{ \pi_k^{(t-1)} f_{\mW}({\bs}_i; {\bSigma^{(t-1)}}_k, n_i-1)}{\sum_{k'=1}^K \pi_{k'}^{(t-1)} f_{\mW}({\bs}_i; {\bSigma^{(t-1)}}_{k'}, n_i-1)}
\end{align*}
and the parameter estimates that maximize the $\mathcal{Q}$-function are found to be
\begin{equation*}
        {\pi}_k^{(t)} = \frac{1}{n}\sum_{i=1}^n \tau_{ik}^{(t)}
        \text{ and    }
       {{\bSigma}}_k^{(t)} = %\frac{\sum_{i=1}^n\tau_{ik}^{(t)} {\bs}_i}{\sum_{i=1}^{n}\tau_{ik}^{(t)}\nu } =
        \frac{\sum_{i=1}^n\tau_{ik}^{(t)} {\bs}_i}{\sum_{i=1}^{n}\tau_{ik}^{(t)}(n_i-1)}.
\end{equation*}
Using this approach, we can modify Algorithm~\ref{alg:Mixture_EM} for the ${\bS}_i \in \mathcal{PD}(p), \text{ } \forall i \in \{1,\dots, n\}$ is shown in Algorithm~\autoref{alg:EM_non_singular_based}.

\begin{algorithm}[H]
\caption{EM Algorithm for Wishart-Based Model}
\label{alg:EM_non_singular_based}
\begin{algorithmic}
\setstretch{1.5}

\STATE \textbf{Input:} Observed data $\mathcal{S} = \{\bs_i\}_{i=1}^{n}$, number of components $K$, convergence threshold $\epsilon > 0$
\STATE \textbf{Initialize:} Parameter set ${\bTheta}^{(0)} = \{\pi_k^{(0)}, \bSigma_k^{(0)}\}_{k=1}^{K}$, set iteration counter $t \gets 0$

\WHILE {$\frac{\mathcal{L}_{\mathcal{W}}({\bTheta}^{(t)}) - \mathcal{L}_{\mathcal{W}}({\bTheta}^{(t-1)})}
{|\mathcal{L}_{\mathcal{W}}({\bTheta}^{(t-1)})|} \geq \epsilon$}

    \STATE \textbf{E-Step:} Compute posterior probabilities (responsibilities):
    
   $ \tau_{ik}^{(t)} \gets \frac{ \pi_k^{(t-1)} f_{\mathcal{W}}(\bs_i \mid \bSigma_k^{(t-1)}, n_i - 1)}
    {\sum_{k'=1}^{K} \pi_{k'}^{(t-1)} f_{\mathcal{W}}(\bs_i \mid \bSigma_{k'}^{(t-1)}, n_i - 1)}
   $ 
    \quad for all $i \in \{1, \dots, n\}$ and $k \in \{1, \dots, K\}$.

    \STATE \textbf{M-Step:} Update parameter estimates:

    \STATE Update mixture weights:
    
  $  \pi_k^{(t)} \gets \frac{1}{n} \sum_{i=1}^{n} \tau_{ik}^{(t)}, \quad \forall k \in \{1, \dots, K\}
  $  

    \STATE Update covariance matrices:
    
    $\bSigma_k^{(t)} \gets \frac{\sum_{i=1}^{n} \tau_{ik}^{(t)} \bs_i}
    {\sum_{i=1}^{n} \tau_{ik}^{(t)} (n_i - 1)}, \quad \forall k \in \{1, \dots, K\}
  $  

    \STATE Increment iteration counter: $t \gets t+1$
    
\ENDWHILE

\STATE \textbf{Output:} Converged estimates $\hat{\bSigma}_k \gets {\bSigma}_k^{(t)}$, $\hat{\pi}_k \gets \pi_k^{(t)}$
\STATE \textbf{return} $\hat{\bSigma}_k, \hat{\pi}_k$

\end{algorithmic}
\end{algorithm}

\subsection{Singular Wishart-based mixture estimation} 
\label{sec: singular_wishart}

Again consider the latent covariance model and the scatter matrix ${\bS}_i = \sum_{j=1}^n({\bX}_{ij} - \bar{{\bX}}_i)({\bX}_{ij} - \bar{{\bX}}_i)^T$. Suppose that $\text{rank}(\bS_i) = a < p \text{ } \forall i \in \{1,\dots, n\}$, where ${\bS}_i \in \mathcal{PSD}(p, a)$. Given the Wishart density, $f_{\mW}(\bA;.)$, is defined on $\mathcal{PD}(p)$, it can quickly be seen that if ${\bA} \in \mathcal{PSD}(p,a)$ then $|{\bA}| = 0$ since some of the eigenvalues of $A$ are zero and thus $f_{\mW}({\bA}; {\bSigma}_k, n_i-1) = 0$. On the other hand, \cite{uhligsingular1994} defined the singular Wishart distribution where ${\bS}_i | Z_i = k \sim \mS\mW({\bSigma}_k, n_i-1)$, where $\mS\mW$ denotes the singular Wishart distribution found in and has the centering matrix ${\bSigma}_k$ with $n_i$ degrees of freedom defined on the space $\mathcal{PSD}(p, a)$. The density of ${\bS}_i$ conditional on $Z_i$ is the density of the singular Wishart distribution
\begin{equation}
   f_{\mS\mW}({\bs}_i; {\bSigma}_k, \nu ) = \left(2^{\frac{\nu p}{2}} |{\bSigma}_k|^{\frac{n_i}{2}} \Gamma_n\left(\frac{\nu }{2}\right)\right)^{-1} \pi^{\frac{\nu ^2-p\nu }{2}} |{\bL}_i|^{\frac{1}{2}(\nu -p-1)}e^{-\frac{1}{2}tr({\bSigma}_k^{-1}{\bs}_i)},
\end{equation}
where ${\bL}_i = \text{diag}(\lambda_1, \lambda_2, \dots, \lambda_a) \in \mathbb{R}^{p \times p}$ and $\lambda_1, \lambda_2, \dots, \lambda_a \in \mathbb{R}$ are the non-zero eigenvalues of ${\bs}_i$.
Similarly to the non-singular Wishart case, the unconditional distribution of ${\bS}_i$ is the finite mixture model
\begin{math}
    f({\bs}; {\bTheta}) = \sum_{i=1}^n\pi_k f_{\mS\mW}(s; {\bSigma}_k, n_i-1).
\end{math}
The likelihood and complete-data \text{log}-likelihood functions are now
% \begin{equation*}
%     \mL({\bTheta}|s^n) = \prod_{i=1}^n \sum_{k=1}^K \pi_k f_{\mS\mW}(s_i; {\bSigma}_k, n_i).
% \end{equation*}
\begin{equation}
    \label{eq:singular_wishart_likelihood}
    \mL_{SW}({\bTheta}|{\bs}^{n}) = \prod_{i=1}^n \sum_{k=1}^K \pi_k f_{\mS\mW}({\bs}; {\bSigma}_k, n_i-1)
\end{equation}
and $\ell_c({\bTheta}|{\bs}^n, Z) = \sum_{i=1}^n \sum_{k=1}^K I(Z_i = k) \log(\pi_k f_{\mS\mW}({\bs}; {\bSigma}_k, n_i-1))$, respectively.
The results from Algorithm~\autoref{alg:Mixture_EM} can be modified and the E-step derived as
\begin{align*}
  \tau_{ik}^{(t)} = P(Z_i = k | {\bs}_i; {\bTheta}^{(t)})
              %&= \frac{ \pi_k^{(t)} f({\bs}_i | Z_i = k; {\bTheta}^{(t)})}{\sum_{k'=1}^K \pi_{k'}^{(t)} f({\bs}_i | Z_i = k'; {\bTheta}^{(t)})} \\
              %&= \frac{ \pi_k^{(t)} f_{\mS\mW}({\bs}_i; {\bSigma}_k, \nu )}{\sum_{k'=1}^K \pi_{k'}^{(t)} f_{\mS\mW}({\bs}_i; {\bSigma}_{k'}, \nu )} \\
              &= \frac{ \pi_k^{(t-1)} f_{\mS\mW}({\bs}_i; {\bSigma^{(t-1)}}_k, n_i-1)}{\sum_{k'=1}^K \pi_{k'}^{(t-1)} f_{\mS\mW}({\bs}_i; {\bSigma^{(t-1)}}_{k'}, n_i-1)}  
\end{align*}
and parameters that maximize the $\mathcal{Q}$-function, the M-step, are found to be
\begin{equation*}
       {\pi}_k^{(t)} = \frac{1}{n}\sum_{i=1}^n \tau_{ik}^{(t)}
        \text{ and    }
       {{\bSigma}}_k^{(t)} %= \frac{\sum_{i=1}^n\tau_{ik}^{(t)} {\bs}_i}{\sum_{i=1}^{n}\tau_{ik}^{(t)}\nu } = 
        \frac{\sum_{i=1}^n\tau_{ik}^{(t)} {\bs}_i}{\sum_{i=1}^{n}\tau_{ik}^{(t)}(n_i-1)}.
\end{equation*}
The EM algorithm for when ${\bS}_i \in \mathcal{PSD}(p,a) \text{ } \forall i \in \{1,\dots, n\}$ is shown in Algorithm~\autoref{alg:EM_singular_based}.

\begin{algorithm}[H]
\caption{EM Algorithm for Singular Wishart-Based Model}
\label{alg:EM_singular_based}
\begin{algorithmic}
\setstretch{1.5}

\STATE \textbf{Input:} Observed data $\mathcal{S} = \{\bs_i\}_{i=1}^{n}$, number of components $K$, convergence threshold $\epsilon > 0$
\STATE \textbf{Initialize:} Parameter set ${\bTheta}^{(0)} = \{\pi_k^{(0)}, \bSigma_k^{(0)}\}_{k=1}^{K}$, set iteration counter $t \gets 0$

\WHILE {$\frac{\mathcal{L}_{\mathcal{SW}}({\bTheta}^{(t)}) - \mathcal{L}_{\mathcal{SW}}({\bTheta}^{(t-1)})}
{|\mathcal{L}_{\mathcal{SW}}({\bTheta}^{(t-1)})|} \geq \epsilon$}

    \STATE \textbf{E-Step:} Compute posterior probabilities (responsibilities):
    
    $\tau_{ik}^{(t)} \gets \frac{ \pi_k^{(t-1)} f_{\mathcal{SW}}(\bs_i \mid \bSigma_k^{(t-1)}, n_i - 1)}
    {\sum_{k'=1}^{K} \pi_{k'}^{(t-1)} f_{\mathcal{SW}}(\bs_i \mid \bSigma_{k'}^{(t-1)}, n_i - 1)}
   $ 
    \quad for all $i \in \{1, \dots, n\}$ and $k \in \{1, \dots, K\}$.

    \STATE \textbf{M-Step:} Update parameter estimates:

    \STATE Update mixture weights:
    
   $ \pi_k^{(t)} \gets \frac{1}{n} \sum_{i=1}^{n} \tau_{ik}^{(t)}, \quad \forall k \in \{1, \dots, K\}
   $ 

    \STATE Update covariance matrices:
    
   $ \bSigma_k^{(t)} \gets \frac{\sum_{i=1}^{n} \tau_{ik}^{(t)} \bs_i}
    {\sum_{i=1}^{n} \tau_{ik}^{(t)} (n_i - 1)}, \quad \forall k \in \{1, \dots, K\}
    $

    \STATE Increment iteration counter: $t \gets t+1$
    
\ENDWHILE

\STATE \textbf{Output:} Converged estimates $\hat{\bSigma}_k \gets {\bSigma}_k^{(t)}$, $\hat{\pi}_k \gets \pi_k^{(t)}$
\STATE \textbf{Return} $\hat{\bSigma}_k, \hat{\pi}_k$

\end{algorithmic}
\end{algorithm}

\subsection{Normal-based mixture estimation} \label{sec: normal}

Note that ${\bX}_{ij} | Z_i = k \sim \mN_p({\bmu}_i, {\bSigma}_k)$ for $i \in \{1,2, \dots, n\}$ and $j \in \{1,2, \dots, n_i\}$. Instead of using the unconditional distribution of the scatter matrix ${\bS}_i = \sum_{j=1}^n({\bX}_{ij} - \bar{{\bX}}_i)({\bX}_{ij} - \bar{{\bX}}_i)^T$, the distribution of ${\bX}_{ij}$ can be used directly to derive the parameter estimates. While it may not be clear that deriving parameter estimates directly using the distribution of ${\bX}_{ij}$ yields an algorithm that clusters both singular and non-singular matrices, as will be seen, this is indeed the case.

The unconditional distribution of ${\bX}_{ij}$ is
\begin{math}
    f({\bx}; {\bPsi}) = \sum_{i=1}^n\pi_k \phi({\bx}; {\bmu}_i, {\bSigma}_k)
\end{math}
which is a finite mixture model where for each class $i$ we have that each mixture component is centered at the mean of the class and where each component of the mixture has a unique covariance matrix. The corresponding likelihood function and complete-data \text{log}-likelihood are then
% \begin{equation}\label{eq:normal_likelihood}
%     \mL_{\mN}({\bPsi}|x^{n}) = \prod_{i=1}^n \sum_{k=1}^K \pi_k \prod_{j=1}^{n_i} \phi(x_{ij}; {\bmu}_i, {\bSigma}_k)
% \end{equation} 
\begin{equation} 
    \label{eq:normal_likelihood}
    \mL_{\mN}({\bPsi}|{\bx}^{n}) = \prod_{i=1}^n \sum_{k=1}^K \pi_k \prod_{j=1}^{n_i} \phi({\bx}_{ij}; {\bmu}_i, {\bSigma}_k)
\end{equation}
and $\ell_c({\bPsi}|{\bx}^n, {\bZ}) = \sum_{i=1}^n \sum_{k=1}^K I(Z_i = k) \log(\pi_k \phi({\bx}_{ij}; {\bmu}_i, {\bSigma}_k))$, with ${\bx}^n = \{ {\bx}_{1}, {\bx}_{2}, \dots, {\bx}_{n} \}$ and ${\bx}_i = \{ {\bx}_{i1}, {\bx}_{i2}, \dots, {\bx}_{in_i} \}$. The $\mathcal{Q}-$function is
\begin{equation*}
    \mathcal{Q}({\bPsi}|{\bx}^n, {\bZ}, {\bPsi}^{(t)}) = E(\ell_c({\bPsi}|{\bx}^n, {\bZ}) | {\bx}^n; {\bPsi}^{(t)}) = \sum_{i=1}^n \sum_{k=1}^K \tau_{ik}^{(t)} \left[ \log(\pi_k) + \sum_{j=1}^{n_i} \log(\phi({\bx}_{ij}; {\bmu}_i, {\bSigma}_k))\right].
\end{equation*}
The E-step and M-step can be derived from the $\mathcal{Q}-$function. For the E-step, the posterior probability is shown to be
\begin{equation*}
    \tau_{ik}^{(t)} %= P(Z_i = k | {\bx}_i; {\bPsi}^{(t-1)}) \\
              = \frac{ \pi_k^{(t)} f({\bx}_i | Z_i = k; {\bPsi}^{(t-1)})}{\sum_{k'=1}^K \pi_{k'}^{(t)} f({\bx}_i | Z_i = k'; {\bPsi}^{(t)})} \\
              = \frac{ \pi_k^{(t)} \prod_{j=1}^{n_i} \phi({\bx}_{ij}; {\bmu}^{(t-1)}_i, {\bSigma}^{(t-1)}_k))}{\sum_{{k'}=1}^K \pi_{k'}^{(t-1)} \prod_{j=1}^{n_i} \phi({\bx}_{ij}; {\bmu}^{(t-1)}_i, {\bSigma}^{(t-1)}_{k'}))}.
\end{equation*}
Next, the parameter estimates that maximize the Q-function are,
\begin{equation*}
       {\pi}_k^{(t)} = \frac{1}{n}\sum_{i=1}^n \tau_{ik}^{(t)} \text{,    }
        {\hat\bmu}_i = \frac{1}{n_i}\sum_{j=1}^{n_i} {\bx}_{ij}
        \text{, and    }
       {{\bSigma}}_k^{(t)} = \frac{\sum_{i=1}^n\tau_{ik}^{(t)} \sum_{j=1}^{n_i} {\bs}_i}{\sum_{i=1}^{n}\tau_{ik}^{(t)}n_i}.
\end{equation*}
Note here that ${\hat\bmu}_i$ does not change across the iterations and ${\bSigma}_k^{(t+1)}$ is taking the weighted average of the scatter matrices of the classes. Thus, allowing for clustering singular and non-singular covariance matrices. The connections between finding parameter estimates using the mixture of Wishart, mixtures of singular Wisharts, and the mixture of normal are illustrated further in Theorem~\autoref{thrm:equality_of_algorithms}. The EM algorithm using the normal-based derivation for when ${\bS}_i \in \mathcal{PSD}(p,a)$ or ${\bS}_i \in \mathcal{PD}(p)$ is shown in Algorithm~\autoref{alg:EM_normal_based}.

\begin{algorithm}[H]
\caption{EM Algorithm for a Normal-Based Model}
\label{alg:EM_normal_based}
\begin{algorithmic}
\setstretch{1.5}

\STATE \textbf{Input:} Observed data $\mathcal{X} = \{\bx_{ij}\}_{i=1, j=1}^{n, n_i}$, number of components $K$, convergence threshold $\epsilon > 0$
\STATE \textbf{Initialize:} $\bTheta^{(0)} = \{\pi_k^{(0)}, \bSigma_k^{(0)}\}_{k=1}^K$, set $t \gets 0$
\STATE Compute class means: $\hat{\bmu}_i \gets \frac{1}{n_i} \sum_{j=1}^{n_i} \bx_{ij}, \quad \forall i \in \{1, \dots, n\}$

\STATE Define the parameter vector: $\bPsi^{(t)} = \{\bTheta^{(t)}, \hat{\bmu}_1, \dots, \hat{\bmu}_n\}$
\WHILE {$\frac{\mathcal{L}_{\mathcal{N}}(\bPsi^{(t)} \mid \mathcal{X}) - \mathcal{L}_{\mathcal{N}}(\bPsi^{(t-1)} \mid \mathcal{X})}{|\mathcal{L}_{\mathcal{N}}(\bPsi^{(t-1)} \mid \mathcal{X})|} \geq \epsilon$}
    \STATE \textbf{E-Step:} Compute posterior probabilities:

   $ \tau_{ik}^{(t)} \gets \frac{ \pi_k^{(t-1)} \prod_{j=1}^{n_i} \phi(\bx_{ij} \mid \hat{\bmu}_i, \bSigma_k^{(t-1)})}
    {\sum_{k'=1}^{K} \pi_{k'}^{(t-1)} \prod_{j=1}^{n_i} \phi(\bx_{ij} \mid \hat{\bmu}_i, \bSigma_{k'}^{(t-1)})}
    $
    \quad for all $i \in \{1, \dots, n\}, k \in \{1, \dots, K\}$.
    \STATE \textbf{M-Step:} Update parameter estimates
    \STATE Update mixture weights:
   $ 
    \pi_k^{(t)} \gets \frac{1}{n} \sum_{i=1}^{n} \tau_{ik}^{(t)}, \quad \forall k \in \{1, \dots, K\}
  $ 
    \STATE Update covariance matrices:
    $
    \bSigma_k^{(t)} \gets \frac{\sum_{i=1}^{n} \tau_{ik}^{(t)} \sum_{j=1}^{n_i} (\bx_{ij} - \hat{\bmu}_i)(\bx_{ij} - \hat{\bmu}_i)^\top}
    {\sum_{i=1}^{n} \tau_{ik}^{(t)} n_i}, \quad \forall k \in \{1, \dots, K\}
    $
    \STATE Increment iteration counter: $t \gets t+1$
 \ENDWHILE
\STATE \textbf{Output:} Converged estimates $\hat{{\bSigma}}_k \gets {\bSigma_k}^{(t)}$, $\hat{\pi}_k \gets \pi_k^{(t)}$
\STATE \textbf{return} $\hat{\bmu}_i, \hat{\bSigma}_k, \hat{\pi}_k$
\end{algorithmic}
\end{algorithm}

\subsection{Theoretical results} 
\label{sec: theory}
\begin{theorem} \label{thrm:equality_of_algorithms}
Given a set of observations ${\bx}^n$ in which ${\bS}_i = \sum_{j=1}^n({\bX}_{ij} - {\bmu}_i)({\bX}_{ij} - {\bmu}_i)^T \in \mathcal{PD}(p) \text{ } \forall i \in \{1,\dots, n\}$, 
% \autoref{alg:EM_normal_based} and \autoref{alg:EM_non_singular_based} are equivalent algorithms in that they result in the same parameter estimates.
the parameter estimates that maximize the likelihoods of the normal-based likelihood (\autoref{eq:normal_likelihood}) and Wishart-based likelihood (\autoref{eq:wishart_likelihood}) are equivalent and furthermore, given a set of initial parameters ${\bTheta}^{(0)} = \{\pi^{(0)}_1, \dots, \pi^{(0)}_K, {\bSigma}^{(0)}_1, \dots, {\bSigma}^{(0)}_K\}$, Algorithms~\autoref{alg:EM_normal_based} and \autoref{alg:EM_non_singular_based} are equivalent algorithms.
Similarly, given a set of observations ${\bx}^n$ in which ${\bS}_i = \sum_{j=1}^n({\bX}_{ij} - {\bmu}_i)({\bX}_{ij} - {\bmu}_i)^T \in \mathcal{PSD}(p,a) \text{, } \forall i \in \{1,\dots, n\}$,
%\autoref{alg:EM_normal_based} and \autoref{alg:EM_non_singular_based} are equivalent algorithms in that they result in the same parameter estimates.
the parameter estimates that maximize the likelihoods of the normal-based likelihood (\autoref{eq:normal_likelihood}) and the singular Wishart-based likelihood (\autoref{eq:singular_wishart_likelihood}) are equivalent and furthermore, given a set of initial parameters ${\bTheta}^{(0)} = \{\pi^{(0)}_1, \dots, \pi^{(0)}_K, {\bSigma}^{(0)}_1, \dots, {\bSigma}^{(0)}_K\}$, Algorithms~\autoref{alg:EM_normal_based} and \autoref{alg:EM_singular_based} of the respective methods are equivalent.
\end{theorem}

\begin{proof}
    First, note the product of normal pdf's with respect to the set ${\bx}^n$  can be rewritten in terms of ${\bs}^n$ by
    \begin{align*}
        \prod_{j=1}^{n_i} \phi({\bx}_{ij}; {\bmu}_i, {\bSigma}_k) &= \prod_{j=1}^{n_i} (2\pi)^{-\frac{p}{2}} |{\bSigma}_k|^{-\frac{1}{2}} e^{-\frac{1}{2} tr\{({\bx}_{ij} - {\bmu}_i)^T {\bSigma}_k^{-1} ({\bx}_{ij} - {\bmu}_i)\}} \\ 
       % &= (2\pi)^{-\frac{n_i p}{2}} |{\bSigma}_k|^{-\frac{n_i}{2}} e^{-\frac{1}{2} \sum_{j=1}^{n_i} tr\{{\bSigma}_k^{-1} ({\bx}_{ij} - {\bmu}_i)({\bx}_{ij} - {\bmu}_i)^T\}} \\
       % &= (2\pi)^{-\frac{n_i p}{2}} |{\bSigma}_k|^{-\frac{n_i}{2}} e^{-\frac{1}{2} tr\{{\bSigma}_k^{-1} \sum_{j=1}^{n_i} ({\bx}_{ij} - {\bmu}_i)({\bx}_{ij} - {\bmu}_i)^T\}} \\
        &= (2\pi)^{-\frac{n_i p}{2}} |{\bSigma}_k|^{-\frac{n_i}{2}} e^{-\frac{1}{2} tr\{{\bSigma}_k^{-1} {\bs}_i\}}.
    \end{align*}
Suppose we have a set of observations ${\bx}^n$, where $n_i \ge p$ for each class. It is then true that ${\bs}_i \in \mathcal{PD}(p)$ and ${\bS}_i | Z_i=k \sim W({\bSigma}_k, n_i)$. This implies that the density of ${\bs}_i$ exists on the space $\mathcal{PD}(p)$.
Assuming that ${\hat\bmu}_i = \frac{1}{n_i}\sum_{j=1}^{n_i} {\bx}_{ij}$ it can be shown
\begin{align*}
    \argmax_{{\bTheta}} \mL_{\mN}({\bTheta}|{\bx}^{n}) &= \argmax_{{\bTheta}} \prod_{i=1}^n \sum_{k=1}^K \pi_k \prod_{j=1}^{n_i} \phi({\bx}_{ij}; {\hat\bmu}_i, {\bSigma}_k) \\
    &=\argmax_{{\bTheta}}\prod_{i=1}^n \sum_{k=1}^K \pi_k (2\pi)^{-\frac{n_i p}{2}} |{\bSigma}_k|^{-\frac{n_i}{2}} e^{-\frac{1}{2} tr\{{\bSigma}_k^{-1} {\bs}_i\}} \\
    &=\argmax_{{\bTheta}}\prod_{i=1}^n (2\pi)^{-\frac{n_i p}{2}} \sum_{k=1}^K \pi_k |{\bSigma}_k|^{-\frac{n_i}{2}} e^{-\frac{1}{2} tr\{{\bSigma}_k^{-1} {\bs}_i\}} \\
    &=\argmax_{{\bTheta}}\prod_{i=1}^n \sum_{k=1}^K \pi_k |{\bSigma}_k|^{-\frac{n_i}{2}} e^{-\frac{1}{2} tr\{{\bSigma}_k^{-1} {\bs}_i\}} \\
   % &=\argmax_{{\bTheta}}\prod_{i=1}^n \left(2^{\frac{n_ip}{2}} \Gamma_p\left(\frac{n_i}{2}\right)\right)^{-1}|{\bs}_i|^{\frac{1}{2}(n-p-1)} \sum_{k=1}^K \pi_k |{\bSigma}_k|^{-\frac{n_i}{2}} e^{-\frac{1}{2} tr\{{\bSigma}_k^{-1} {\bs}_i\}} \\
    &=\argmax_{{\bTheta}}\prod_{i=1}^n \sum_{k=1}^K \pi_k \left(2^{\frac{n_ip}{2}} \Gamma_p\left(\frac{n_i}{2}\right)\right)^{-1}|{\bs}_i|^{\frac{1}{2}(n-p-1)} |{\bSigma}_k|^{-\frac{n_i}{2}} e^{-\frac{1}{2} tr\{{\bSigma}_k^{-1} {\bs}_i\}} \\
    &=\argmax_{{\bTheta}}\prod_{i=1}^n \sum_{k=1}^K \pi_k f_{\mW}({\bs}_i; {\bSigma}_k, n_i) = \argmax_{{\bTheta}}\mL_{\mW}({\bTheta}; {\bs}^n)
\end{align*}
and thus $\argmax_{{\bTheta}} \mL_{\mN}({\bTheta}|{\bx}^{n}) = \argmax_{{\bTheta}}\mL_{\mW}({\bTheta}; {\bs}^n)$.
Consider the E-step of the normal-based derivation \autoref{alg:EM_normal_based}. It can quickly be seen that coefficients not dependent on $k$ can be factored out, yielding
\begin{align*}
    \tau_{ik}^{(t)} &= \frac{ \pi_k^{(t)} \prod_{j=1}^{n_i} \phi({\bx}_{ij}; {\bmu}_i, {\bSigma}_k))}{\sum_{{k'}=1}^K \pi_{k'}^{(t)} \prod_{j=1}^{n_i} \phi({\bx}_{ij}; {\bmu}_i, {\bSigma}_{k'}))} \\
    &= \frac{ \pi_k^{(t)} (2\pi)^{-\frac{n_i p}{2}} |{\bSigma}_k|^{-\frac{n_i}{2}} e^{-\frac{1}{2} tr\{{\bSigma}_k^{-1} {\bs}_i\}}}{\sum_{{k'}=1}^K \pi_{k'}^{(t)} (2\pi)^{-\frac{n_i p}{2}} |{\bSigma}_{k^{'}}|^{-\frac{n_i}{2}} e^{-\frac{1}{2} tr\{{\bSigma}_{k^{'}}^{-1} {\bs}_i\}}} \\
    &= \frac{ \pi_k^{(t)} |{\bSigma}_k|^{-\frac{n_i}{2}} e^{-\frac{1}{2} tr\{{\bSigma}_k^{-1} {\bs}_i\}}}{\sum_{{k'}=1}^K \pi_{k'}^{(t)} |{\bSigma}_{k^{'}}|^{-\frac{n_i}{2}} e^{-\frac{1}{2} tr\{{\bSigma}_{k^{'}}^{-1} {\bs}_i\}}}. 
\end{align*}
Similarly, for the E-step of Wishart-based derivation 
\autoref{alg:EM_non_singular_based} we have
\begin{align*}
    \tau_{ik}^{(t)} &= \frac{ \pi_k^{(t)} f_{\mW}({\bs}_{ii}; {\bSigma}_k))}{\sum_{{k'}=1}^K \pi_{k'}^{(t)} f_{\mW}({\bs}_{i}; {\bSigma}_{k'}))} \\
    &= \frac{ \pi_k^{(t)} \left(2^{\frac{n_ip}{2}} \Gamma_p\left(\frac{n_i}{2}\right)\right)^{-1}|{\bs}_i|^{\frac{1}{2}(n-p-1)} |{\bSigma}_k|^{-\frac{n_i}{2}} e^{-\frac{1}{2} tr\{{\bSigma}_k^{-1} {\bs}_i\}}}{\sum_{{k'}=1}^K \pi_{k'}^{(t)} \left(2^{\frac{n_ip}{2}} \Gamma_p\left(\frac{n_i}{2}\right)\right)^{-1}|{\bs}_i|^{\frac{1}{2}(n-p-1)} |{\bSigma}_{k^{'}}|^{-\frac{n_i}{2}} e^{-\frac{1}{2} tr\{{\bSigma}_{k^{'}}^{-1} {\bs}_i\}}} \\
    &= \frac{ \pi_k^{(t)} |{\bSigma}_k|^{-\frac{n_i}{2}} e^{-\frac{1}{2} tr\{{\bSigma}_k^{-1} {\bs}_i\}}}{\sum_{{k'}=1}^K \pi_{k'}^{(t)} |{\bSigma}_k|^{-\frac{n_i}{2}} e^{-\frac{1}{2} tr\{{\bSigma}_k^{-1} {\bs}_i\}}}. 
\end{align*}
Therefore, Algorithm~\autoref{alg:EM_normal_based} is equivalent to Algorithm~\autoref{alg:EM_non_singular_based} given the same initial values ${\bTheta}^{(0)}$ and yield the same parameter estimates. In addition, for normal-based derivations, we have
%Suppose we have a set of observations ${\bx}^n$ where $n_1 = n_2 = \dots n_n = a < p$ for each class. Then each ${\bs}_i \in \mathcal{PSD}(p,a)$ which implies that the density of ${\bs}_i$ exists on the space $\mathcal{PSD}(p,a)$. Also we have that ${\bS}_i | Z_i=k \sim SW({\bSigma}_k, n_i)$. Again assuming that ${\bmu}_i = \frac{1}{n_i}\sum_{j=1}^{n_i} {\bx}_{ij}$, can then be seen that
\begin{align*}
    \argmax_{{\bTheta}} \mL_{\mN}({\bTheta}|{\bx}^{n}) &=\argmax_{{\bTheta}}\prod_{i=1}^n \sum_{k=1}^K \pi_k |{\bSigma}_k|^{-\frac{n_i}{2}} e^{-\frac{1}{2} tr\{{\bSigma}_k^{-1} {\bs}_i\}} \\
    &=\argmax_{{\bTheta}}\prod_{i=1}^n \left(2^{\frac{n_ip}{2}} \Gamma_p\left(\frac{n_i}{2}\right)\right)^{-1} \pi^{\frac{n_i^2-pn_i}{2}} |{\bL}_i|^{\frac{1}{2}(n-p-1)} \sum_{k=1}^K \pi_k |{\bSigma}_k|^{-\frac{n_i}{2}} e^{-\frac{1}{2} tr\{{\bSigma}_k^{-1} {\bs}_i\}} \\
    &=\argmax_{{\bTheta}}\prod_{i=1}^n \sum_{k=1}^K \pi_k \left(2^{\frac{n_ip}{2}} \Gamma_p\left(\frac{n_i}{2}\right)\right)^{-1} \pi^{\frac{n_i^2-pn_i}{2}} |{\bL}_i|^{\frac{1}{2}(n-p-1)} |{\bSigma}_k|^{-\frac{n_i}{2}} e^{-\frac{1}{2} tr\{{\bSigma}_k^{-1} {\bs}_i\}} \\
    &=\argmax_{{\bTheta}}\prod_{i=1}^n \sum_{k=1}^K \pi_k f_{\mS\mW}({\bs}_i; {\bSigma}_k, n_i) = \argmax_{{\bTheta}}L_{\mS\mW}({\bTheta}; {\bs}^n)
\end{align*}
and thus $\argmax_{{\bTheta}} \mL_{\mN}({\bTheta}|{\bx}^{n}) = \argmax_{{\bTheta}}L_{\mS\mW}({\bTheta}; {\bs}^n)$.
% \begin{align*}
%     \tau_{ik}^{(t)} &= \frac{ \pi_k^{(t)} f_{\mW}(s_{i}; {\bSigma}_k))}{\sum_{{k'}=1}^K \pi_{k'}^{(t)} f_{\mW}(s_{i}; {\bSigma}_{k'}))} \\
%     &= \frac{ \pi_k^{(t)} \left(2^{\frac{n_ip}{2}} \Gamma_p\left(\frac{n_i}{2}\right)\right)^{-1}|s_i|^{\frac{1}{2}(n-p-1)} |\Sigma_k|^{-\frac{n_i}{2}} e^{-\frac{1}{2} tr\{\Sigma_k^{-1} s_i\}}}{\sum_{{k'}=1}^K \pi_{k'}^{(t)} \left(2^{\frac{n_ip}{2}} \Gamma_p\left(\frac{n_i}{2}\right)\right)^{-1}|s_i|^{\frac{1}{2}(n-p-1)} |\Sigma_{k^{'}}|^{-\frac{n_i}{2}} e^{-\frac{1}{2} tr\{\Sigma_{k^{'}}^{-1} s_i\}}} \\
%     &= \frac{ \pi_k^{(t)} |\Sigma_k|^{-\frac{n_i}{2}} e^{-\frac{1}{2} tr\{\Sigma_k^{-1} s_i\}}}{\sum_{{k'}=1}^K \pi_{k'}^{(t)} |\Sigma_k|^{-\frac{n_i}{2}} e^{-\frac{1}{2} tr\{\Sigma_k^{-1} s_i\}}}. 
% \end{align*}
The E-step for \autoref{alg:EM_non_singular_based} can then be show to be equal to the E-step for \autoref{alg:EM_normal_based} by
\begin{align*}
    \tau_{ik}^{(t)} &= \frac{ \pi_k^{(t)} f_{\mS\mW}({\bs}_{ii}; {\bSigma}_k))}{\sum_{{k'}=1}^K \pi_{k'}^{(t)} f_{\mS\mW}({\bs}_{i}; {\bSigma}_{k'}))} \\
    &= \frac{ \pi_k^{(t)} \left(2^{\frac{n_ip}{2}} \Gamma_p\left(\frac{n_i}{2}\right)\right)^{-1} \pi^{\frac{n_i^2-pn_i}{2}} |{\bL}_i|^{\frac{1}{2}(n-p-1)} |{\bSigma}_k|^{-\frac{n_i}{2}} e^{-\frac{1}{2} tr\{{\bSigma}_k^{-1} {\bs}_i\}}}{\sum_{{k'}=1}^K \pi_{k'}^{(t)} \left(2^{\frac{n_ip}{2}} \Gamma_p\left(\frac{n_i}{2}\right)\right)^{-1} \pi^{\frac{n_i^2-pn_i}{2}} |{\bL}_i|^{\frac{1}{2}(n-p-1)} |{\bSigma}_{k^{'}}|^{-\frac{n_i}{2}} e^{-\frac{1}{2} tr\{{\bSigma}_{k^{'}}^{-1} {\bs}_i\}}} \\
    &= \frac{ \pi_k^{(t)} |{\bSigma}_k|^{-\frac{n_i}{2}} e^{-\frac{1}{2} tr\{{\bSigma}_k^{-1} {\bs}_i\}}}{\sum_{{k'}=1}^K \pi_{k'}^{(t)} |{\bSigma}_k|^{-\frac{n_i}{2}} e^{-\frac{1}{2} tr\{{\bSigma}_k^{-1} {\bs}_i\}}}. 
\end{align*}
and therefor \autoref{alg:EM_normal_based} is equivalent to \autoref{alg:EM_singular_based} given the same initial values ${\bTheta}^{(0)}$ and yield the same parameter estimates.
% \qedsymbol{}
\end{proof}

\paragraph{\textbf{Neyman-Scott Problem}}
The Neyman-Scott Problem \cite{jiang_large_2022} is a well-known example where the maximum likelihood estimate of a distribution yields an asymptotically inconsistent estimate. Following from \cite{jiang_large_2022} but for the multivariate version, consider ${\bY}_{ij} \sim N_p({\bmu}_i, {\bSigma})$ for $i \in \{1,2, \dots, n\}$ and $j \in \{1,2, \dots, m\}$. Then the MLEs can be shown to be $\hat{{\bmu}}_i = \frac{1}{m} \sum_{j=1}^m {\bY}_{ij}$ and $\hat{{\bSigma}} = \frac{1}{nm} \sum_{i=1}^n \sum_{j=1}^m ({\bY}_{ij} - \hat{{\bmu}}_i) ({\bY}_{ij} - \hat{{\bmu}}_i)^T$. It can then be seen that $\hat{{\bSigma}} \xrightarrow{p} \frac{m-1}{m} {\bSigma}$ as $n \rightarrow \infty$. In other words, even though the number of classes is going to infinity, the estimate of the shared covariance matrix is still biased. It is not until the number of observations within a class $m$, tends to infinity that the estimate is asymptotically unbiased though this paper is concerned with situations $m$ is small and fixed. One remedy is to ``correct" for the bias with by $\tilde{{\bSigma}} = \frac{m}{m-1} \hat{{\bSigma}}$ where now $\tilde{{\bSigma}} \xrightarrow{p} {\bSigma}$ as $n \rightarrow \infty$. This paper encounters the mixture version of the Neyman-Scott Problem and is illustrated in \autoref{thrm:asym_biased}.

\begin{theorem} \label{thrm:asym_biased}
    The maximum likelihood estimates for ${\bSigma}_k$ that maximize the likelihood from \autoref{eq:normal_likelihood} via \autoref{alg:EM_normal_based} are asymptotically biased as $n \rightarrow \infty$. Particularly
    \begin{equation*}
        \hat{{\bSigma}}_k^{*} \xrightarrow{p} \left(\frac{\sum_{i=1}^n\tau_{ik}^{*} (n_i - 1)}{\sum_{i=1}^{n}\tau_{ik}^{*}n_i} \right) {\bSigma}_k \text{ as } n \rightarrow \infty,
    \end{equation*}
    where $\tau_{ik}^{*}$ is the posterior probability under the parameters that maximize the likelihood.
    % and asymptotically unbiased if $n_i \rightarrow \infty$ for at least one class $i$ that is parameterized by $\Sigma_k$.
\end{theorem}

\begin{proof}
    Consider the MLE estimates ${\bTheta}^{*} = \{\pi^*_1, \dots, \pi^*_K, {\bmu}^*_1, \dots, {\bmu}^*_n, {\bSigma}^*_1, \dots, {\bSigma}^*_K\}$ that maximize the likelihood of \autoref{alg:EM_normal_based}. Using ${\bTheta}^{*}$ as the parameters of the current iteration $(t)$ of \autoref{alg:EM_normal_based} such that ${\bTheta}^{(t)}={\bTheta}^{*}$, the $\mathcal{Q}-$function is
    \begin{equation*}
        \mathcal{Q}({\bTheta}|{\bx}^n, Z, {\bTheta}^{(t)}) = \sum_{i=1}^n \sum_{k=1}^K \tau_{ik}^{(t)} \left[ \log(\pi_k) + \sum_{j=1}^{n_i} \log(\phi({\bx}_{ij}; {\bmu}_i, {\bSigma}_k))\right].
    \end{equation*}
    Taking the derivative of the $\mathcal{Q}-$function with respect to ${\bSigma}_k$ we have
    \begin{equation*}
        \frac{\partial Q}{\partial {\bSigma}_k} = \frac{\partial}{\partial {\bSigma}_k} \sum_{i=1}^n \tau_{ik}^{(t)} \sum_{j=1}^{n_i} \log(\phi({\bx}_{ij}; {\bmu}_i, {\bSigma}_k)).
    \end{equation*}
    This is equivalent to to assuming ${\bX}_{ij} \sim N({\bmu}_i,  {\bSigma}_k)$ and finding the estimate of ${\bSigma}_k$ that maximizes the weighted log-likelihood \cite{wang2004asymptotic} where the weights are the set $\{\tau_{ik}^{(t)} \}_{i \in \{1, \dots, n\}, k \in \{1, \dots, K\}}$. The parameter estimate at convergence is found to be
    \begin{equation*}
        \hat{{\bSigma}}_k = \frac{\sum_{i=1}^n\tau_{ik} \sum_{j=1}^{n_i} ({\bx}_{ij} - \hat{{\bmu}}_i)({\bx}_{ij} - \hat{{\bmu_i}})^T}{\sum_{i=1}^{n}\tau_{ik}n_i} = \frac{\sum_{i=1}^n\tau_{ik}{\bs}_i}{\sum_{i=1}^{n}\tau_{ik}n_i}
    \end{equation*}
    This is now a weighted version of the Neyman-Scott problem.
    We then have that ${\bs}_i \xrightarrow{p} (n_i - 1) {\bSigma}_k$ as $n \rightarrow \infty$ which implies that at convergence the parameter estimate for the covariance matrix 
    \begin{equation*}
        \hat{{\bSigma}}_k \xrightarrow{p} \frac{\sum_{i=1}^n\tau_{ik} (n_i - 1) {\bSigma}_k}{\sum_{i=1}^{n}\tau_{ik}n_i} \text{ as } n \rightarrow \infty.
    \end{equation*}
    % Since Algorithm~\autoref{alg:EM_normal_based} converges to MLEs \cite{wu1983convergence} and the algorithm was assumed to start at ${\bTheta}^*$, we have that ${\bTheta}^{(t+1)}={\bTheta}^{*}$ which implies
    % \begin{equation*}
    %     \hat{{\bSigma}}_k \xrightarrow{p} \left(\frac{\sum_{i=1}^n\tau_{ik} (n_i - 1)}{\sum_{i=1}^{n}\tau_{ik}n_i} \right) {\bSigma}_k \text{ as } n \rightarrow \infty.
    % \end{equation*}
    Thus, the maximum likelihood estimates of \autoref{eq:normal_likelihood} are asymptotically inconsistent.
\end{proof}
\begin{corollary}
Given the results from \autoref{thrm:asym_biased}, a new parameter estimate for ${\bSigma}_k$ can be defined which is unbiased similarly to that found in the Neyman-Scott problem. Particularly, define the parameter estimate to be
\begin{equation*}
        \tilde{{\bSigma}}_k = \left(\frac{\sum_{i=1}^n\tau_{ik} n_i}{\sum_{i=1}^{n}\tau_{ik}(n_i - 1)} \right) \hat{{\bSigma}}_k
\end{equation*}
where we now have that $\tilde{{\bSigma}}_k \xrightarrow{p} {\bSigma}_k$ as $n \rightarrow \infty$. Thus Algorithm~\autoref{alg:EM_normal_based} can be modified to account for the adjusted estimates as given in Algorithm~\autoref{alg:EM_normal_based_adj}.

\begin{algorithm}[H]
\caption{EM Algorithm for a normal-Based Model with adjusted estimate}
\label{alg:EM_normal_based_adj}
\begin{algorithmic}
\setstretch{1.5}

\STATE \textbf{Input:} Observed data $\mathcal{X} = \{\bx_{ij}\}_{i=1, j=1}^{n, n_i}$, number of components $K$, convergence threshold $\epsilon > 0$
\STATE \textbf{Initialize:} $\bTheta^{(0)} = \{\pi_k^{(0)}, \bSigma_k^{(0)}\}_{k=1}^K$, set $t \gets 0$
\STATE Compute class means: $\hat{\bmu}_i \gets \frac{1}{n_i} \sum_{j=1}^{n_i} \bx_{ij}, \quad \forall i \in \{1, \dots, n\}$

\STATE Define the parameter vector: $\bPsi^{(t)} = \{\bTheta^{(t)}, \hat{\bmu}_1, \dots, \hat{\bmu}_n\}$
\WHILE {$\frac{\mathcal{L}_{\mathcal{N}}(\bPsi^{(t)} \mid \mathcal{X}) - \mathcal{L}_{\mathcal{N}}(\bPsi^{(t-1)} \mid \mathcal{X})}{|\mathcal{L}_{\mathcal{N}}(\bPsi^{(t-1)} \mid \mathcal{X})|} \geq \epsilon$}
    \STATE \textbf{E-Step:} Compute posterior probabilities:

   $ \tau_{ik}^{(t)} \gets \frac{ \pi_k^{(t-1)} \prod_{j=1}^{n_i} \phi(\bx_{ij} \mid \hat{\bmu}_i, \bSigma_k^{(t-1)})}
    {\sum_{k'=1}^{K} \pi_{k'}^{(t-1)} \prod_{j=1}^{n_i} \phi(\bx_{ij} \mid \hat{\bmu}_i, \bSigma_{k'}^{(t-1)})}
    $
    \quad for all $i \in \{1, \dots, n\}, k \in \{1, \dots, K\}$.
    \STATE \textbf{M-Step:} Update parameter estimates
    \STATE Update mixture weights:
   $ 
    \pi_k^{(t)} \gets \frac{1}{n} \sum_{i=1}^{n} \tau_{ik}^{(t)}, \quad \forall k \in \{1, \dots, K\}
  $ 
    \STATE Update covariance matrices:
    $
    \bSigma_k^{(t)} \gets \frac{\sum_{i=1}^{n} \tau_{ik}^{(t)} \sum_{j=1}^{n_i} (\bx_{ij} - \hat{\bmu}_i)(\bx_{ij} - \hat{\bmu}_i)^\top}
    {\sum_{i=1}^{n} \tau_{ik}^{(t)} n_i}, \quad \forall k \in \{1, \dots, K\}
    $
    \STATE Increment iteration counter: $t \gets t+1$
 \ENDWHILE
\STATE \textbf{Output:} Converged estimates $\hat{{\bSigma}}_k \gets \left(\frac{\sum_{i=1}^n\tau_{ik}^{(t+1)} n_i}{\sum_{i=1}^{n}\tau_{ik}^{(t+1)}(n_i - 1)} \right){\bSigma_k}^{(t)}$, $\hat{\pi}_k \gets \pi_k^{(t)}$
\STATE \textbf{return} $\hat{\bmu}_i, \hat{\bSigma}_k, \hat{\pi}_k$
\end{algorithmic}
\end{algorithm}

\end{corollary}

\begin{proposition} \label{thrm:lda_qda}

If $K=1$ then the latent covariance model is equivalent to the LDA model and \autoref{alg:EM_normal_based} yields equivalent parameter estimates to that of LDA. If $K=n$ then the latent covariance model is equivalent to the QDA model and \autoref{alg:EM_normal_based} yields equivalent parameter estimates to that of QDA.

\end{proposition}

\begin{theorem} \label{thrm:bayes_rules}
    Given a new observation ${\by}$ and the previous observations ${\bx}^n$, the Bayes decision rule for LCDA is
    \begin{equation*}
        \argmax_{i \in \{1, \dots, n\}} \sum_{k=1}^K \tau_{ik} \phi(\B{y}; {\bmu}_i, {\bSigma}_k)
    \end{equation*}
    where $\tau_{ik} = P(Z_i = k | {\bx}_i)$.
\end{theorem}

\begin{proof}
\small
Conciser a new observation ${\by}$ which was sampled from an unknown class $C$. The rule that maximizes the zero-one loss function for classification problems in known to be 
\\
$\argmax_{i \in \{1, \dots, n\}} P(C=i | {\bx}^n, {\by})$. Using the law of iterative expectations and Bayes theorem we see that
\begin{align*}
    \argmax_{i \in \{1, \dots, n\}} P(C=i | {\bx}^n, {\by}) &= \argmax_{i \in \{1, \dots, n\}} E(I(C=i) | {\bx}^n, {\by}) \\
    &=\argmax_{i \in \{1, \dots, n\}} E(E(I(C=i) | {\bx}^n, {\by}, Z_i) | {\bx}^n, {\by}) \\
    %&=\argmax_{i \in \{1, \dots, n\}} E(P(C=i | {\bx}^n, {\by}, Z_i) | {\bx}^n, {\by}) \\
    &=\argmax_{i \in \{1, \dots, n\}} E\left(\sum_{k=1}^K I(Z_i=k)P(C=i | {\bx}^n, {\by}, Z_i=k) | {\bx}^n, {\by}\right) \\
   % &=\argmax_{i \in \{1, \dots, n\}} \sum_{k=1}^K P(Z_i=k | {\bx}^n, {\by})P(C=i | {\bx}^n, {\by}, Z_i=k) \\
    &=\argmax_{i \in \{1, \dots, n\}} \sum_{k=1}^K P(Z_i=k | {\bx}_i, {\by})P(C=i | {\by}, Z_i=k) \\
    &=\argmax_{i \in \{1, \dots, n\}} \sum_{k=1}^K \frac{\pi_k f({\bx}_i, {\by} | Z_i=k)}{f({\bx}_i, {\by})} \frac{P(C=i) f({\by} | Z_i=k, C=i)}{f({\by})} \\
    %&=\argmax_{i \in \{1, \dots, n\}} \sum_{k=1}^K \frac{\pi_k f({\bx}_i, | Z_i=k)f({\by})}{f({\bx}_i)f({\by})} \frac{P(C=i) f({\by} | Z_i=k , C=i)}{f({\by})} \\
    &=\argmax_{i \in \{1, \dots, n\}} \sum_{k=1}^K \frac{\pi_k f({\bx}_i, | Z_i=k)}{f({\bx}_i)} P(C=i) f({\by} | Z_i=k , C=i) \\
   % &=\argmax_{i \in \{1, \dots, n\}} \sum_{k=1}^K P(Z_i=k | {\bx}_i) P(C=i) f({\by} | Z_i=k , C=i) \\
    &=\argmax_{i \in \{1, \dots, n\}} \sum_{k=1}^K \tau_{ik} f({\by} | Z_i=k , C=i) \\
    &=\argmax_{i \in \{1, \dots, n\}} \sum_{k=1}^K \tau_{ik} \phi({\by} | {\bmu}_i, {\bSigma}_k)
    % \argmax_{i \in \{1, \dots, n\}} P(C=i | {\bx}^n, {\by}) &= \argmax_{i \in \{1, \dots, n\}} \frac{f({\bx}_i, {\by} | C=i)P(C=i)}{f({\bx}_i, {\by})} \\
    % &= \argmax_{i \in \{1, \dots, n\}} f({\bx}^n, {\by} | C=i) \\
    % &= \argmax_{i \in \{1, \dots, n\}} \sum_{k=1}^K \pi_k f({\bx}_i, {\by} | C=i, Z_i=k) \\
    % &= \argmax_{i \in \{1, \dots, n\}} \sum_{k=1}^K \pi_k f({\bx}_i | Z_i=k) f({\by} | C=i, Z_i=k) \\
    % &= \argmax_{i \in \{1, \dots, n\}} \frac{1}{\sum_{k'=1}^K \pi_{k'} f({\bx}_i | Z_i=k')} \sum_{k=1}^K \pi_k f({\bx}_i | Z_i=k) f({\by} | C=i, Z_i=k) \\
    % &= \argmax_{i \in \{1, \dots, n\}} \sum_{k=1}^K \frac{\pi_k f({\bx}_i | Z_i=k)}{\sum_{k'=1}^K \pi_{k'} f({\bx}_i | Z_i=k')} f({\by} | C=i, Z_i=k) \\
    % &= \argmax_{i \in \{1, \dots, n\}} \sum_{k=1}^K \tau_{ik} \phi({\by} | {\bmu}_i, {\bSigma}_k)
\end{align*}
\end{proof}

\subsection{Classification}
The main goal of clustering covariance matrices in this paper is to improve classification accuracy as compared to LDA and QDA by relaxing the strength of the assumptions being made. The typical approach of LDA and QDA for classifying a new observation into one of the given classes is to use the Bayes decision rule. For LDA the assumption is made that ${\bX}_{ij} \sim N({\bmu}_i, {\bSigma})$ for $i \in \{1,2, \dots, n\}$ and $j \in \{1,2, \dots, n_i\}$. A new observation ${\bx}$, is then assigned to a class by $\argmax_i \phi({\bx}; {\bmu}_i, {\bSigma})$ assuming equal prior probability for each class. For QDA the assumption is made that ${\bX}_{ij} \sim N({\bmu}_i, {\bSigma}_i)$ for $i \in \{1,2, \dots, n\}$ and $j \in \{1,2, \dots, n_i\}$. A new observation $\B{x}$, is then assigned to a class by $\argmax_i \phi({\bx}; {\bmu}_i, {\bSigma}))$ assuming equal prior probability for each class. Using the assumptions outlined in this paper for use in classification is referred to Latent Covariance Discriminant Analysis (LCDA) where ${\bX}_{ij} | Z_i = k \sim N_p({\bmu}_i, {\bSigma}_k)$ for $i \in \{1,2, \dots, n\}$ and $j \in \{1,2, \dots, n_i\}$ for $Z_i \sim Categorical(\pi_1, \pi_2, \dots, \pi_K)$. The optimal Bayes rule for LCDA is given in \autoref{thrm:bayes_rules}. This is ofcourse assuming the true parameters of the model are known and thus in this paper we use the plug-in optimal Bayes decision rule using either the MLEs or the adjusted estimates.

\subsection{Initialization and Model Selection} \label{sec:init, conv, selection}

Since the EM algorithm is an iterative process, it must be seeded with initial parameter estimates. While there are numerous different methods for initialization of the EM algorithm \cite{melnykovinitializing2012, michaeleffective2016} some of which are stochastic and others are deterministic. For the scope of this paper, a deterministic method is used but it is noted that further study is needed for initializing this algorithm. More specifically, hierarchical clustering is used. The algorithm is outlined in \ref{alg:initialization}.

\begin{algorithm}[H]
\caption{Hierarchical Clustering-Based Initialization}
\label{alg:initialization}
\begin{algorithmic}
\setstretch{1.5}

\STATE \textbf{Input:} Data $\mathcal{X} = \{\bx_{ij}\}_{j=1}^{n_i}$ for $i = 1, \dots, n$, number of clusters $K$

\STATE \textbf{Step 1: Compute Scatter Matrices}
\STATE Compute within-group scatter matrices:

$\bs_i \gets \sum_{j=1}^{n_i}(\bx_{ij} - \bar{\bx}_i)(\bx_{ij} - \bar{\bx}_i)^T, \quad \forall i \in \{1, \dots, n\}$

where $\bar{\bx}_i = \frac{1}{n_i} \sum_{j=1}^{n_i} \bx_{ij}$.

\STATE \textbf{Step 2: Define Similarity Function}
\STATE Define a matrix similarity measure:

$d(M_1, M_2) = \|\left(M_1^{\frac{1}{2}}\right) - \left(M_2^{\frac{1}{2}}\right)\|$
where $M^{\frac{1}{2}} = V \Lambda^{1/2} V^T$ is the matrix square root obtained from the eigenvalue decomposition $M = V \Lambda V^T$.

\STATE \textbf{Step 3: Hierarchical Clustering}
\STATE Perform hierarchical clustering using Ward's linkage:
\STATE Construct hierarchical tree $T$ on $\{\bs_1, \dots, \bs_n\}$ using distance function $d$.
\STATE Cut tree $T$ to obtain $K$ clusters and assign cluster labels $C = \{C_1, \dots, C_n\}$.

\STATE \textbf{Step 4: Compute Initial Parameter Estimates}
\STATE Estimate initial mixture weights:

$\pi_k \gets \frac{1}{n} \sum_{i=1}^{n} I(C_i = k), \quad \forall k \in \{1, \dots, K\}$

\STATE Estimate initial covariance matrices:

$\bSigma_k \gets \frac{\sum_{i=1}^{n} I(C_i = k) \bs_i}{\sum_{i=1}^{n} I(C_i = k) n_i}, \quad \forall k \in \{1, \dots, K\}$

\STATE \textbf{Output:} Initial estimates  $ \bTheta^{(0)}=\{\pi_k,\bSigma_k\}$
\STATE \textbf{return} $\bTheta^{(0)}$

\end{algorithmic}
\end{algorithm}

In order to choose the number of latent covariance matrices $K$, an information criterion is used. Particularly the Bayesian information criterion (BIC) \cite{schwarz1978estimating} where $BIC = m ln(n) - 2 ln(\hat{\mL})$ where $m$ is the number of parameters in the models and $\hat{\mL}$ is the likelihood given the parameter estimates.

\section{Empirical study}
 \label{Simulated Data}
 
A simulation study is performed in order to evaluate the performance of the clustering solutions, model selection, and classification accuracy of LCDA as compared to LDA and QDA. To do this, $25$ random models are generated under each of the settings; Dimensionality $p \in \{4, 8, 12\}$, number of latent covariance matrices $K \in \{2, 4, 6\}$, number of classes $n \in \{100, 200, 300\}$ and number of within class observations $n_i^* \in \{p/2, [p/2, 2p], 2p\}$. Hence, if $n_i^* = p/2$, then for each class, $p/2$ observations are sampled, if $n_i^* =[p/2, 2p]$ then for each class $i$ a number $n_i$ is sampled uniformly from the interval $[p/2 2p]$ at which point $n_i$ observations are sampled from the $i$th class. Finally, if $n_i^* = 2p$, then for each class $i$, $2p$ observations are sampled from each class. Note that in the scheme outlined, there will be cases when each class has a singular covariance estimate, when each class has a non-singular covariance estimate, and when some classes have a singular covariance estimate while others have non-singular estimates. The mean of each class was uniformly sampled from a hypercube of the respective dimensions.

\subsection{Clustering Accuracy}
\begin{figure}%
    \centering
   % \subfloat[\centering ARI over the different model settings]
    {\includegraphics[width=7.8cm]{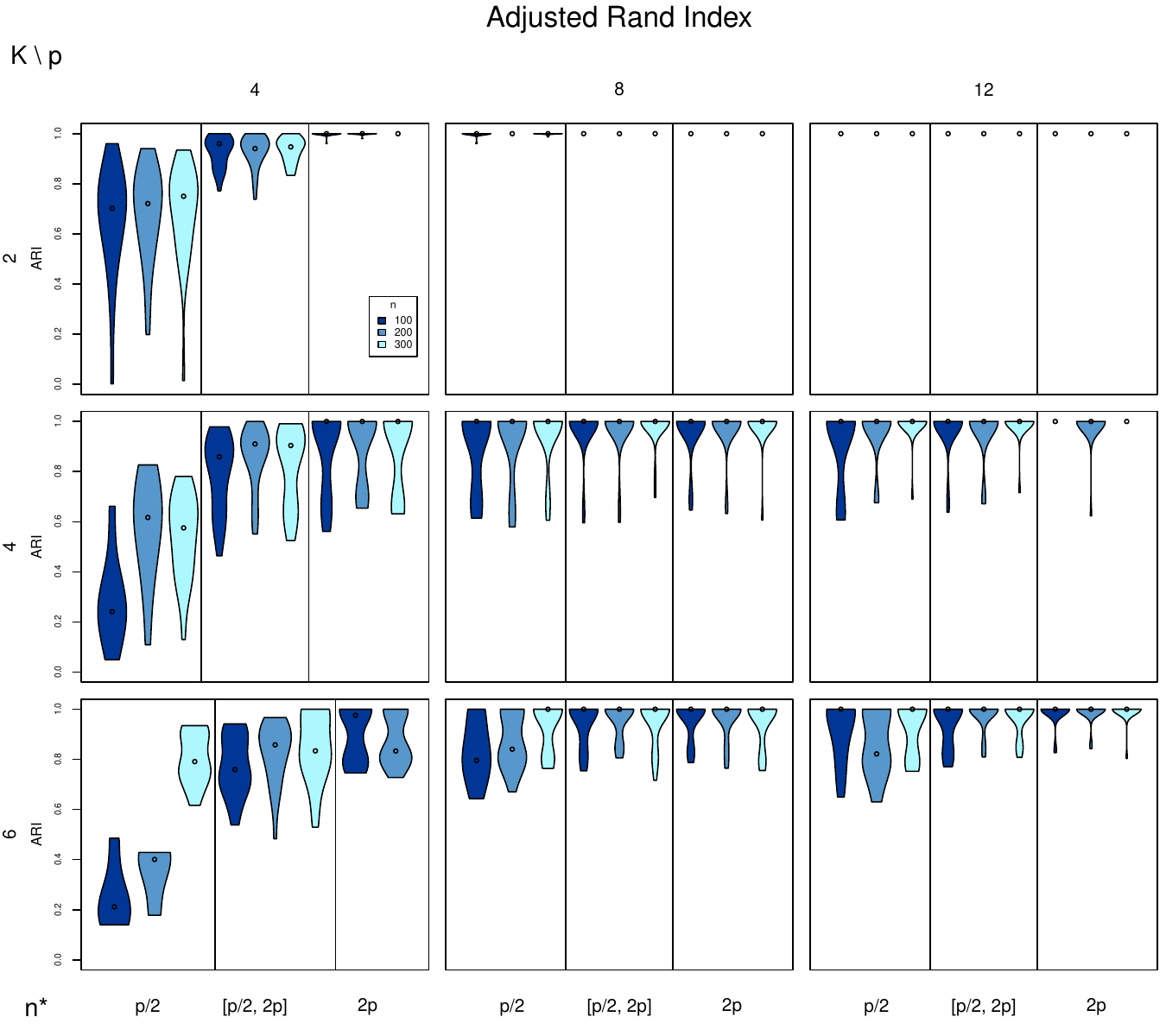} }%
    \hspace{0.1cm}
    %\subfloat[\centering Model selection over the different model settings using BIC]
    {{\includegraphics[width=7.8cm]{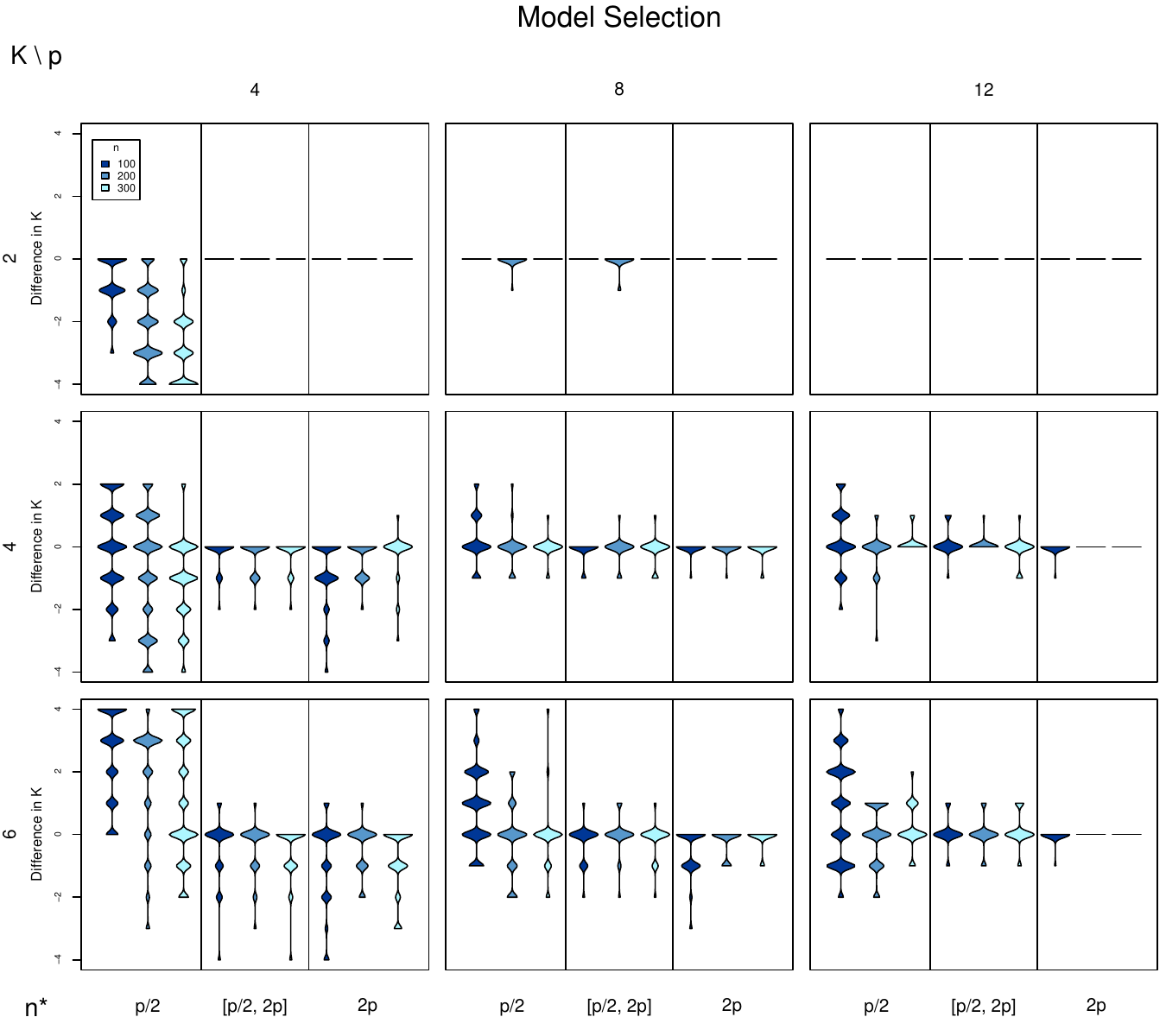} }}%
    \caption{ARI and model selection via BIC for combination of hyperparameters ($K, p, n_i, n$).}%
    \label{fig:ari_model_selection}%
\end{figure}

The first performance metric of interest is the Adjusted Rand Index (ARI) \cite{hubert1985comparing}, which is a measure of similarity between the clustering solution found and the true clustering solution. An ARI value of $1$ indicates a perfect clustering solution. We also compared the true $K$ with one selected using BIC and computed the difference $K -\hat K$. The results across the different settings can be seen in \autoref{fig:ari_model_selection}. From subplot~\autoref{fig:ari_model_selection}(a), we can see that the algorithm performs well at clustering covariance matrices. Some expected results can be seen where as the number of classes increases, the number of dimensions increases, or the number of samples within a class increases, it becomes easier to cluster the covariance matrices, and thus, the ARI value increases. Similarly, as expected, as the number of latent covariance matrices $K$ increases, it becomes more difficult to cluster the covariance matrices, and thus, the ARI decreases.

%\subsection{Model Selection}
Next, the performance of the model selection criterion for choosing the number of latent covariance matrices $K$, which in this case is BIC, is investigated. The results can be seen in subplot \autoref{fig:ari_model_selection}(b), where the value of interest is the difference between the true value of $K$ and the value picked using BIC. Some intuitive results can be seen: when the number of classes increases, the number of dimensions increases, or the number of samples within a class increases, the model selection problem becomes easier. Across the plots, it can be seen that when the number of samples within a class is low, {\it i.e.,} $n^*=p/2$, the model selection problem is particularly more challenging. In this case, though, we see that as the number of classes increases, the model selection performance also increases.

\subsection{Accuracy compared to LDA and QDA}

\begin{figure}%
    \centering
    %\subfloat[\centering Odds ratio of LCDA over LDA. The red horizontal line plotted at one indicated which model has better performance]
    {{\includegraphics[width=7.8cm]{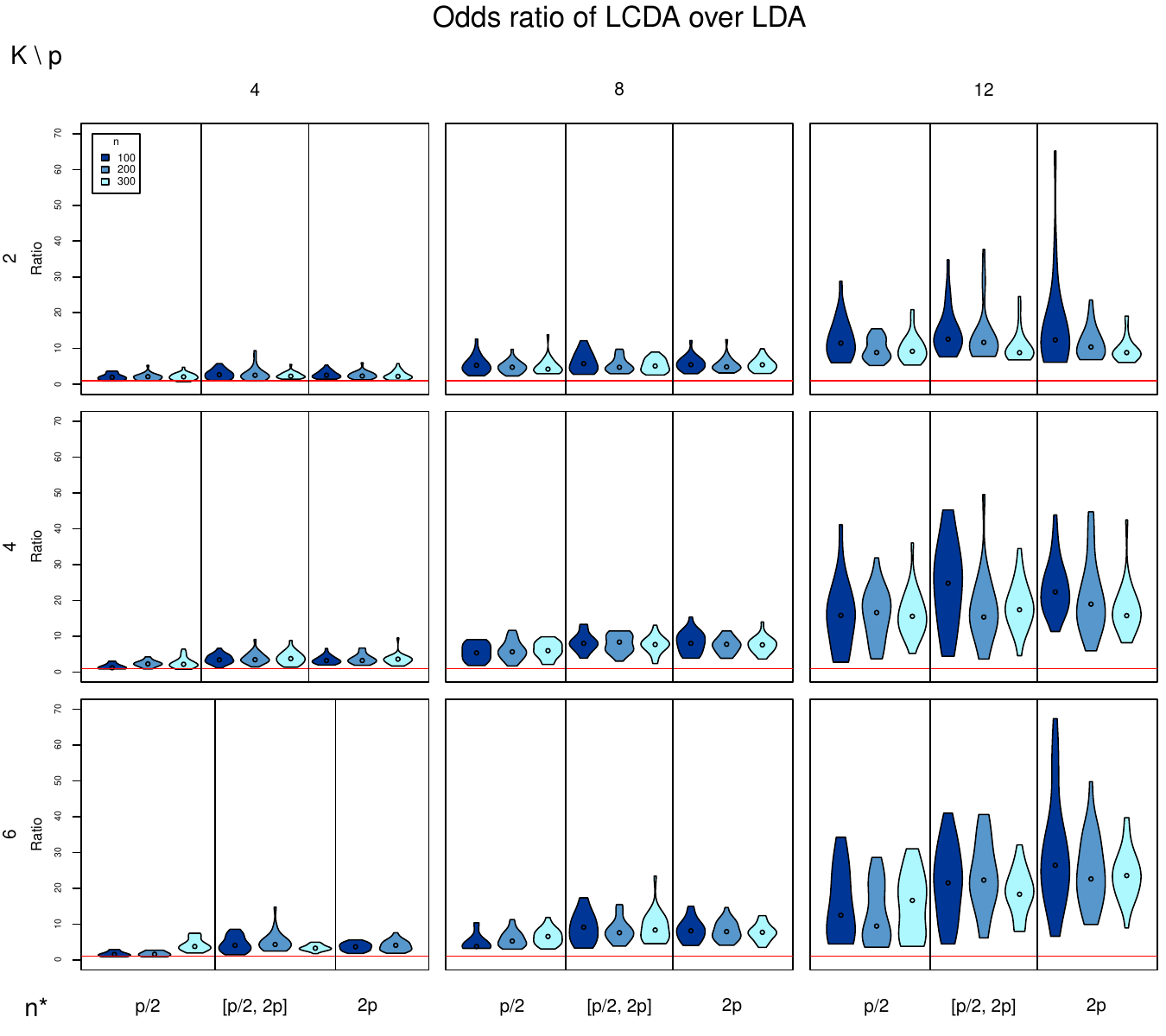} }}%
    \hspace{0.1cm}
    %\subfloat[\centering Odds ratio of LCDA over QDA. The red horizontal line plotted at one indicated which model has better performance]
    {{\includegraphics[width=7.8cm]{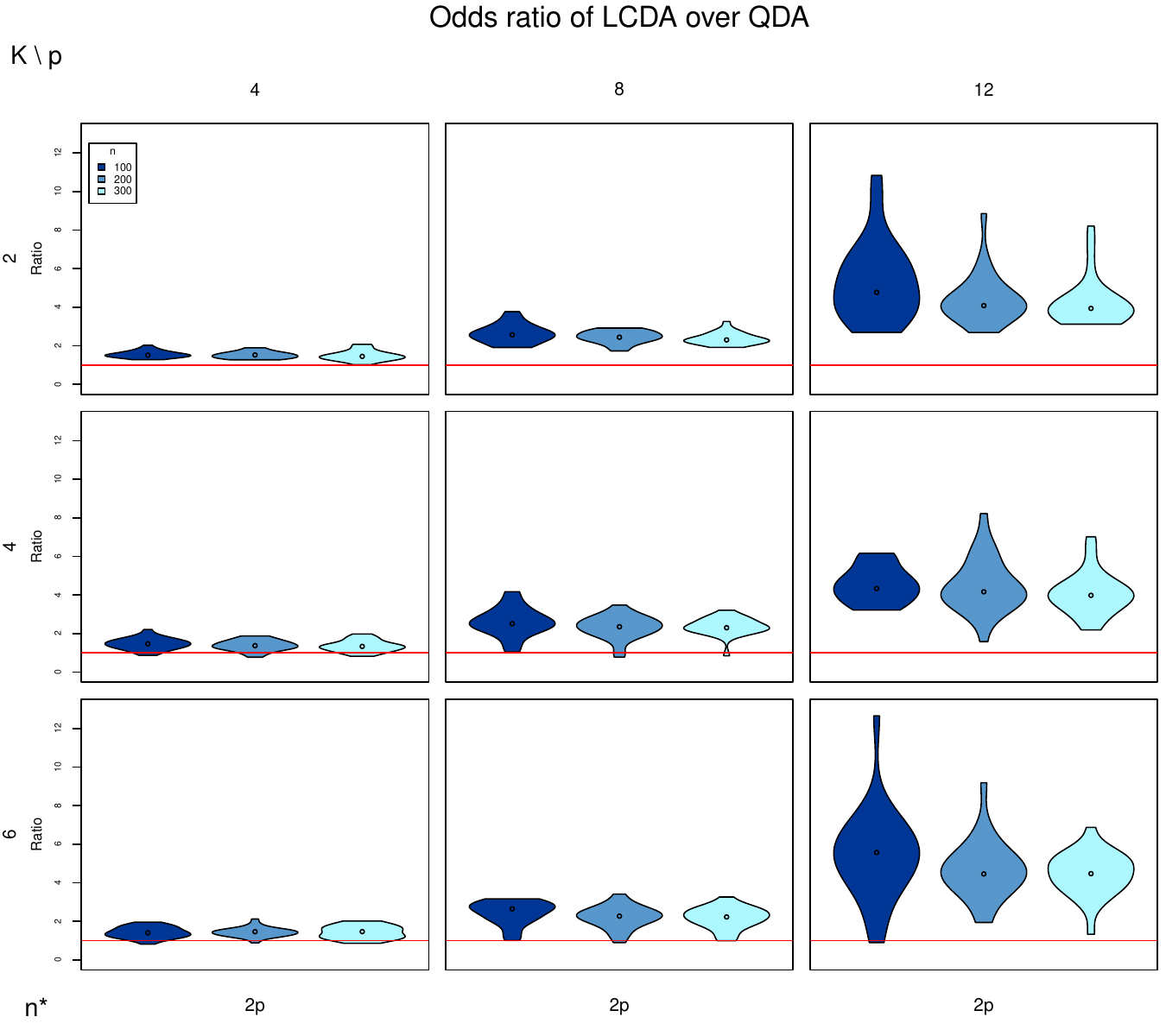} }}%
    \caption{Comparing the accuracy of LCDA, LDA and QDA}%
    \label{fig:lda_qda}%
\end{figure}

While the data in this simulation study was generated under the assumptions of LCDA and thus LCDA is expected to perform better than LDA, it is still interesting to observe ``how much" better it performs across the different model settings. On the other hand, whenever it is possible to fit QDA and get a positive definite matrix, one might expect QDA to outperform LCDA since it is more flexible. Let $\mA_m \in [0,1]$ be the accuracy from model $m$. Suppose we wish to compare the accuracy of two methods relative to each other. Directly comparing $\frac{\mA_{LCDA}}{\mA_{LDA}}$ for example may not be very informative as $\mA_m$ is bounded by 1. Thus we instead compare the odds ratio where the odds is $\frac{\mA_m}{1-\mA_m} \in [0, \infty)$ and the odds ratio of models $m_1$ over $m_2$ is $\frac{\mA_{m_1} / (1 - \mA_{m_1})}{\mA_{m_2} / (1 - \mA_{m_2})}$. Thus, a ratio greater than one suggests the model in the numerator is performing better, whereas a ratio less than one suggests the model in the denominator is performing better. For LCDA, the adjusted estimates are used as the plug-in estimates from the optimal decision rule for each comparison. In addition, accuracy is based on out-of-sample predictions of classes. From \autoref{fig:lda_qda}, it can be seen that in every setting, LCDA performance is better than LDA and QDA, as expected. When comparing LCDA to LDA, we can see that as $K$ increases, the difference in the performance between the models also increases. This does not necessarily seem to be the case for QDA. The more notable observation taken from \autoref{fig:lda_qda} is the difference in performance between LCDA and both LDA and QDA as the dimensions of the data increase. In the case of $p=12$, there is around a $20\times$ boost in performance from LCDA over LDA and a $4\times$ boost from LCDA over QDA. This illustrates the problem of severely breaking assumptions in the LDA and QDA in high-dimensional data.

\subsection{MLE vs REML estimates of the latent covariance matrices}
\begin{figure}%
    \centering
    {\includegraphics[width=14cm]{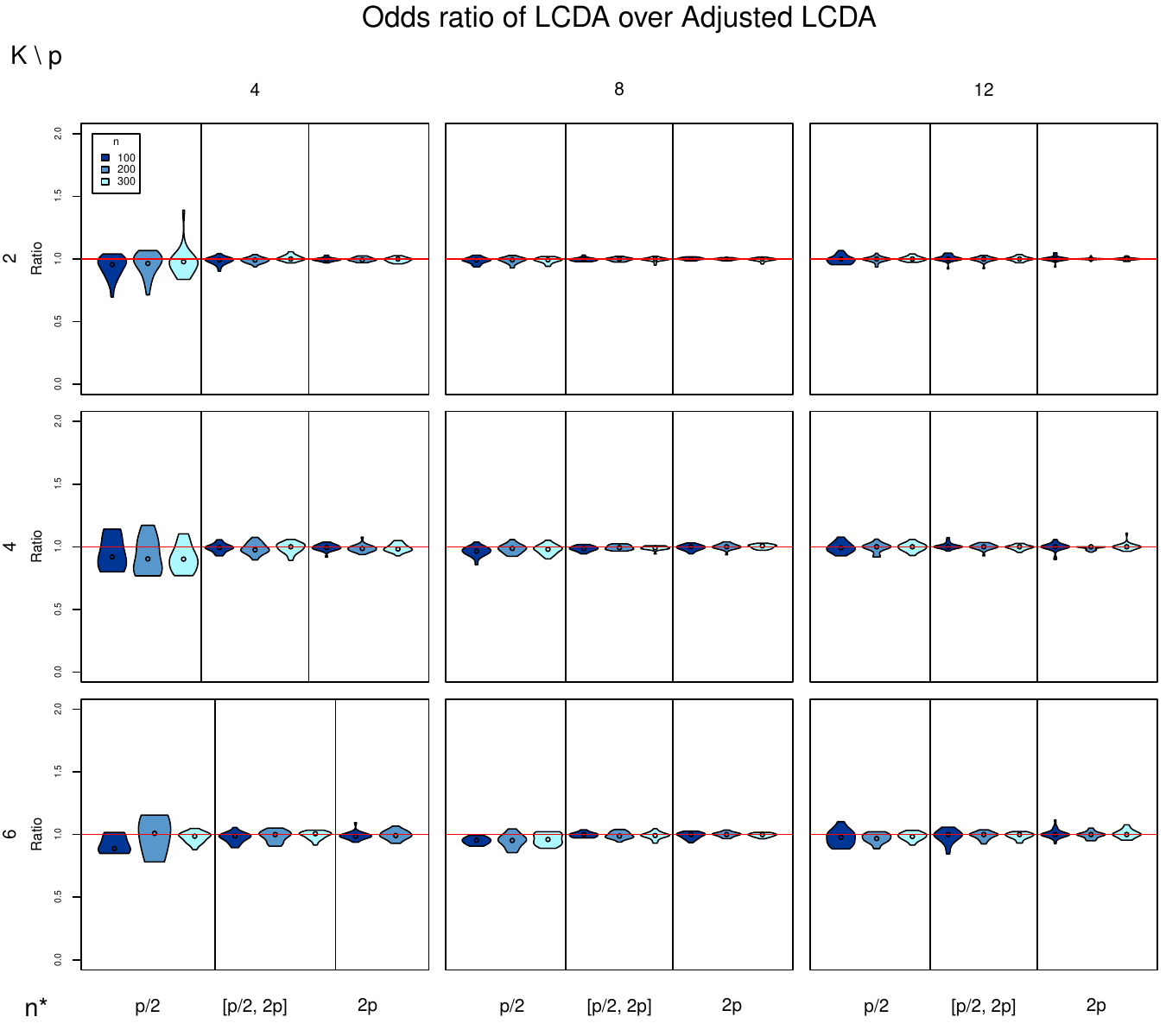} }%
    \caption{Odds ratio of LCDA over the adjusted LCDA using the REML estimates. The red horizontal line plotted at one indicated which model has better performance}%
    \label{fig:ldca_adjlcda}%
\end{figure}

Further empirical study is done to investigate the consistency of theoretical results and its effects on the few-shot classification accuracy. While using the adjusted estimates (REML) of the latent covariance matrices, $\hat\bSigma_k$, had the desirable consistency over the MLEs, comparing these two estimates in terms of accuracy might be interesting when using them as plug-in estimates of the optimal decision rule. Hence, these two approaches were compared within each of the hyperparameter combinations ($K, p, n_i, n$). Similarly to the above, here we also look at the out-of-sample accuracy of each approach. The odds ratio results are seen in \autoref{fig:ldca_adjlcda}. In general, we see a slight increase in performance using the adjusted estimates over the MLEs, though this difference is mitigated as $p$ increases, as $n$ increases, and as $n^*$ increases.

\section{Application}
\label{Real Data}

The methods proposed in this paper are implemented on two different datasets. The first is a glass fragment dataset, where the goal is to classify a new fragment of unknown origin back to its source. The second dataset is a Kannada handwriting dataset, where the goal is to classify an image of handwritten Kannada characters and symbols. In both examples, the proposed LCDA method is compared to LDA and, when possible, with QDA.

\subsection{Zadora Glass Dataset}
\begin{figure}[t]%
    \centering
    {\includegraphics[width=0.90\textwidth]{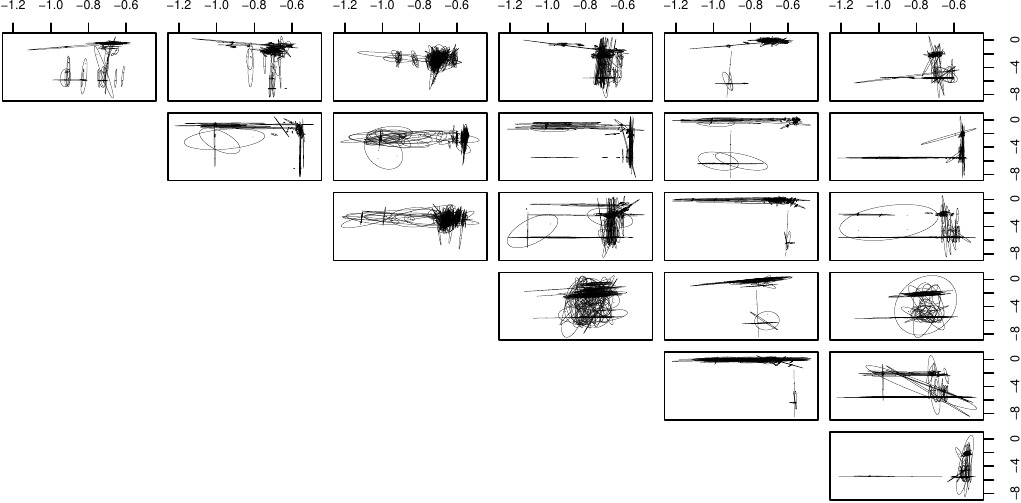} }%
    \caption{The $95\%$ confidence ellipses of each class centered at the mean and using sample covariance matrices.}%
    \label{fig:glass_data_qda}%
\end{figure}

The Zadora glass dataset \cite{aitkentwo-level2007} consists of $200$ objects where the glass came from, with $4$ fragments from each object where each fragment was measured $3$ times. The measurements taken from each fragment are the amounts of the trace elements Sodium, Magnesium, Aluminium, Silicon, Potassium, Calcium, Iron, and Oxygen. The dataset provided is the log of the ratio of the trace elements over Oxygen, yielding a seven-dimensional dataset. Since each fragment is measured $3$ times, constituting technical replicates, the mean is taken over these $3$ measurements. Thus, this dataset has $200$ classes in a $7$ dimensional space where each class has $4$ observations. Taking the sample mean and sample covariance estimates and plotting the $95\%$ confidence ellipse of each class can be seen in \autoref{fig:glass_data_qda}. While the two ellipses can be plotted is two dimensions at a time it must be noted that the full covariance estimate for each class in this case is singular since the number of observations per class is less than the number of dimensions. Thus QDA is not possible in this situation. 

\begin{figure}%
    \centering
    %\subfloat[\centering BIC across different values of $K$ for Zadora's glass dataset]
    {{\includegraphics[width=0.48\textwidth]{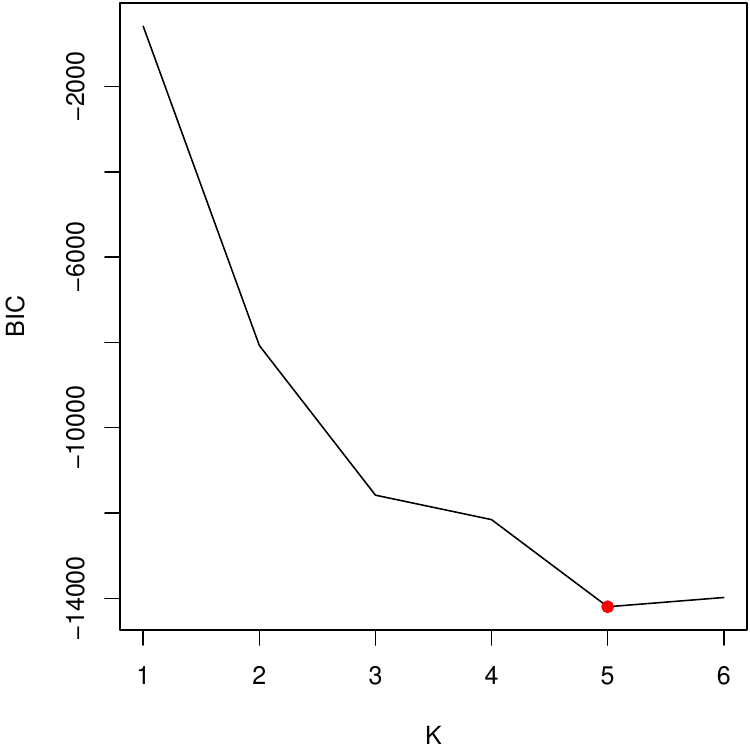} }}%
    \hspace{0.1cm}
    %\subfloat[\centering Proportion of classes from the glass dataset with the respective accuracy using leave one out cross validation]
    {{\includegraphics[width=0.48\textwidth]{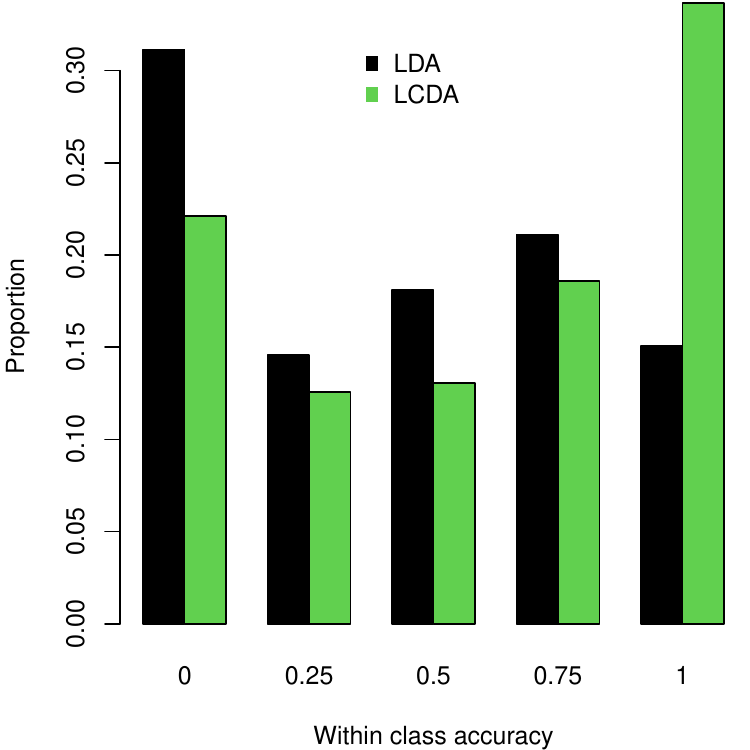} }}%
    \caption{Analysis of LCDA on the glass dataset}%
    \label{fig:bic_results}%
\end{figure}

Model selection is first performed using BIC to apply LCDA to the glass dataset. From \autoref{fig:bic_results}(a), we can see that BIC selects $5$ components in this situation. To compare the performance of LCDA and LDA for the glass dataset, a leave-one-out cross-validation (LOOCV) is performed. Since there are $4$ observations per class, we can get an accuracy for each respective class from LOOCV where the accuracy can be zero where all four were misclassified, or 0.25 where three out of four are misclassified, and so on. So, for each class, we can have LOOCV rates of $\mM = \{0, 0.25, 0.5, 0.75, 1\}$. Then, we can find the proportion of total classes with accuracy rates out of the 200 classes for each method. The results are presented as a barplot in \autoref{fig:bic_results}(b). LCDA strongly outperforms LDA as LCDA had fewer classes with an LOOCV within class accuracy of $0, 0.25$, and $0.75$, and far more classes with LOOCV within class accuracy of $1$.

\begin{figure}[H]%
    \centering
    {\includegraphics[width=14cm]{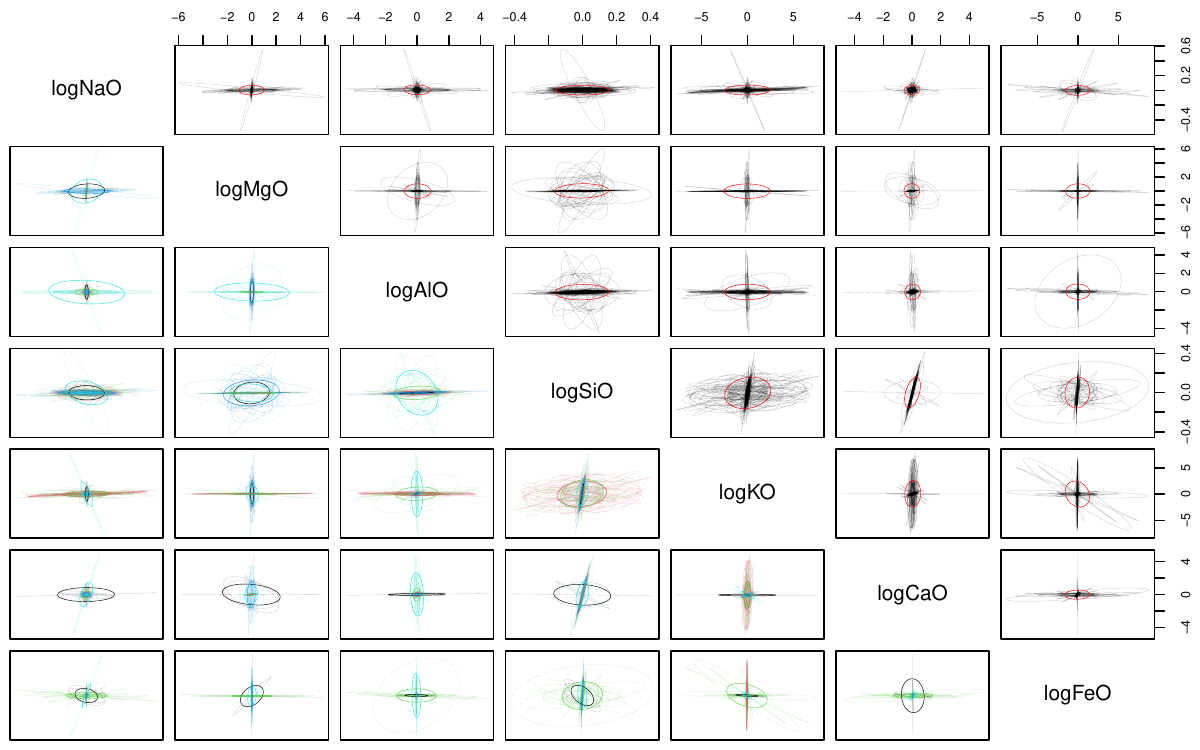} }%
    \caption{Pairs plot of the showing the $95\%$ confidence ellipses of each class centered at the zero. The bottom triangular plots show the clustering solution and covariance estimate from LCDA. The upper triangular plots show the parameter estimate of the covariance estimate from LDA (red ellipse) and centered scatter matrices (grey ellipses).}%
    \label{fig:glass_clustering}%
\end{figure}

A visualization of the clustering solution found by the proposed method can be seen in \autoref{fig:glass_clustering}. The bottom triangular plots show the results from LCDA. The faded ellipses are the centered $95\%$ confidence ellipses for each class, where the color represents the cluster. The bold ellipses represent the $95\%$ confidence ellipses of the estimated covariance matrix for that respective cluster. Similarly is the upper triangular plots to the right, where the bold red ellipse represents the $95\%$ confidence ellipse of the estimated covariance matrix under the assumptions of LDA. This plot illustrates the strong assumption of LDA when there are clearly clusters of covariance matrices that are able to be captured better under the assumptions of LCDA. For this dataset, the accuracy from LOOCV was $44\%$ for LDA and increased to $57\%$ for LCDA.

\subsection{Kannada handwriting dataset}

\begin{figure}[h]%
    \centering
    %\subfloat[\centering Class 1]
    {{\includegraphics[width=4.5cm]{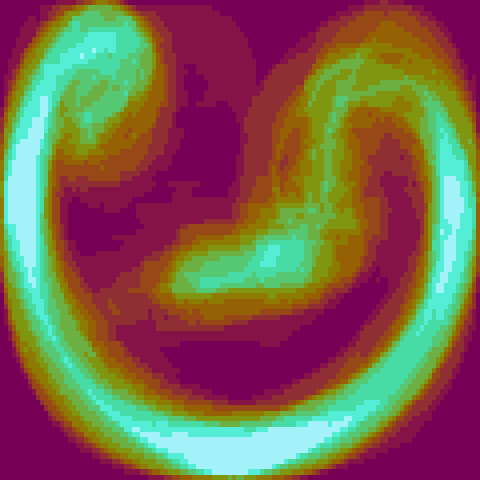} }}%
    \hspace{0.1cm}
    %\subfloat[\centering Class 2]
    {{\includegraphics[width=4.5cm]{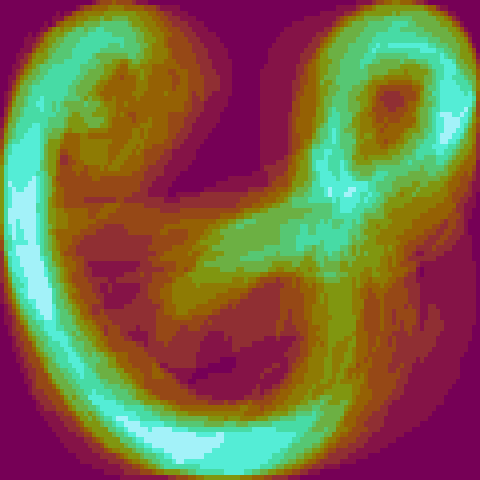} }}%
    \hspace{0.1cm}
    %\subfloat[\centering Class 3]
    {{\includegraphics[width=4.5cm]{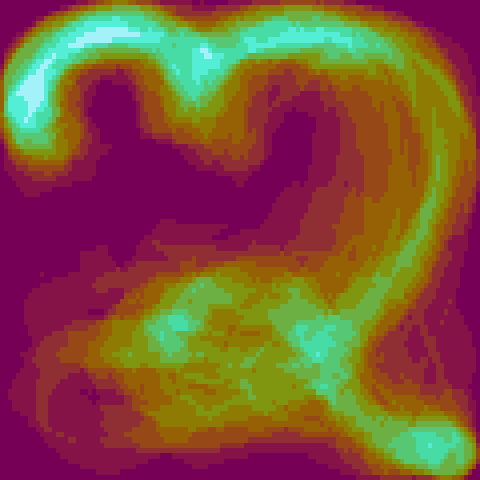} }}%
    \caption{Images showing the averages of the first 3 classes in the Kannada dataset}
    \label{fig:kannada_examples}
\end{figure}

This dataset from \cite{deCampos09} consists of images of handwritten Kannada characters and symbols. There are $657$ classes in this dataset, consisting of $49$ unique characters and visually unique combinations of consonants and vowels. Each class has $25$ respective images. Examples from this dataset are seen in \autoref{fig:kannada_examples}. While each class has many more observations than in the glass dataset, there are many more classes here and in a higher dimensional space. Since the observations are images in this setting, feature extraction methods must first be applied. There are many options for this, and comparing the performance of LCDA to LDA is not necessarily dependent on this choice as the proposed method assumes the observations already live in the Euclidean space. This paper used the histogram of orientated gradients \cite{dalal2005histograms} method using a $3 \times 3$ grid with $6$ bins per cell. This yields a $54$ dimensional space, which was then reduced further to $30$ using principal component analysis (PCA) \cite{wold1987principal}. Thus, we have $657$ classes with $25$ observations per class and in a $30$-dimensional space. 

\begin{figure}%
    \centering
   % \subfloat[\centering  LCDA vs LDA as the number of samples per class decreases]
    {{\includegraphics[width=7.8cm]{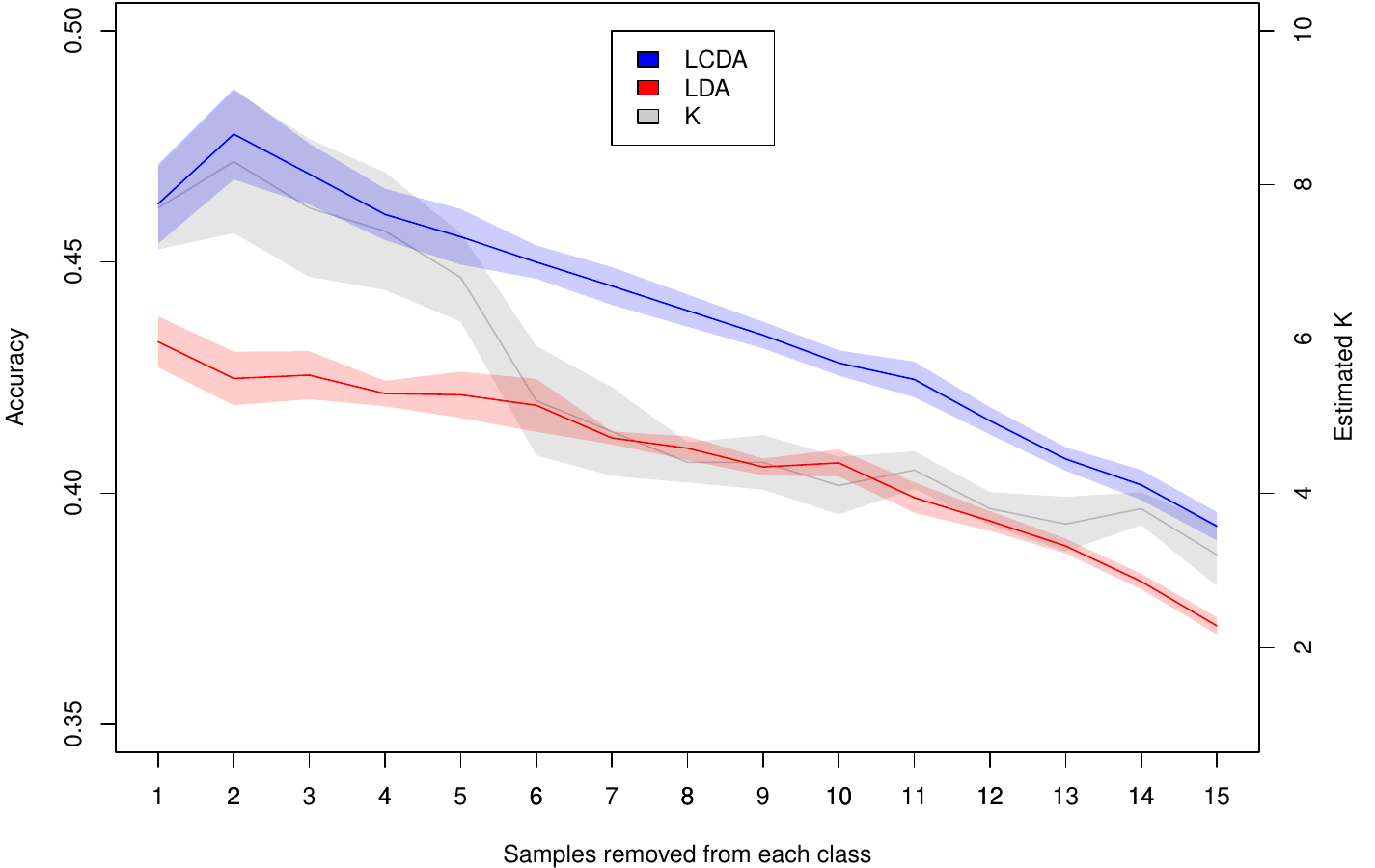} }}%
    \hspace{0.0cm}
   % \subfloat[\centering LCDA, LDA, and QDA as the dimensions of the data increases]
    {{\includegraphics[width=7.8cm]{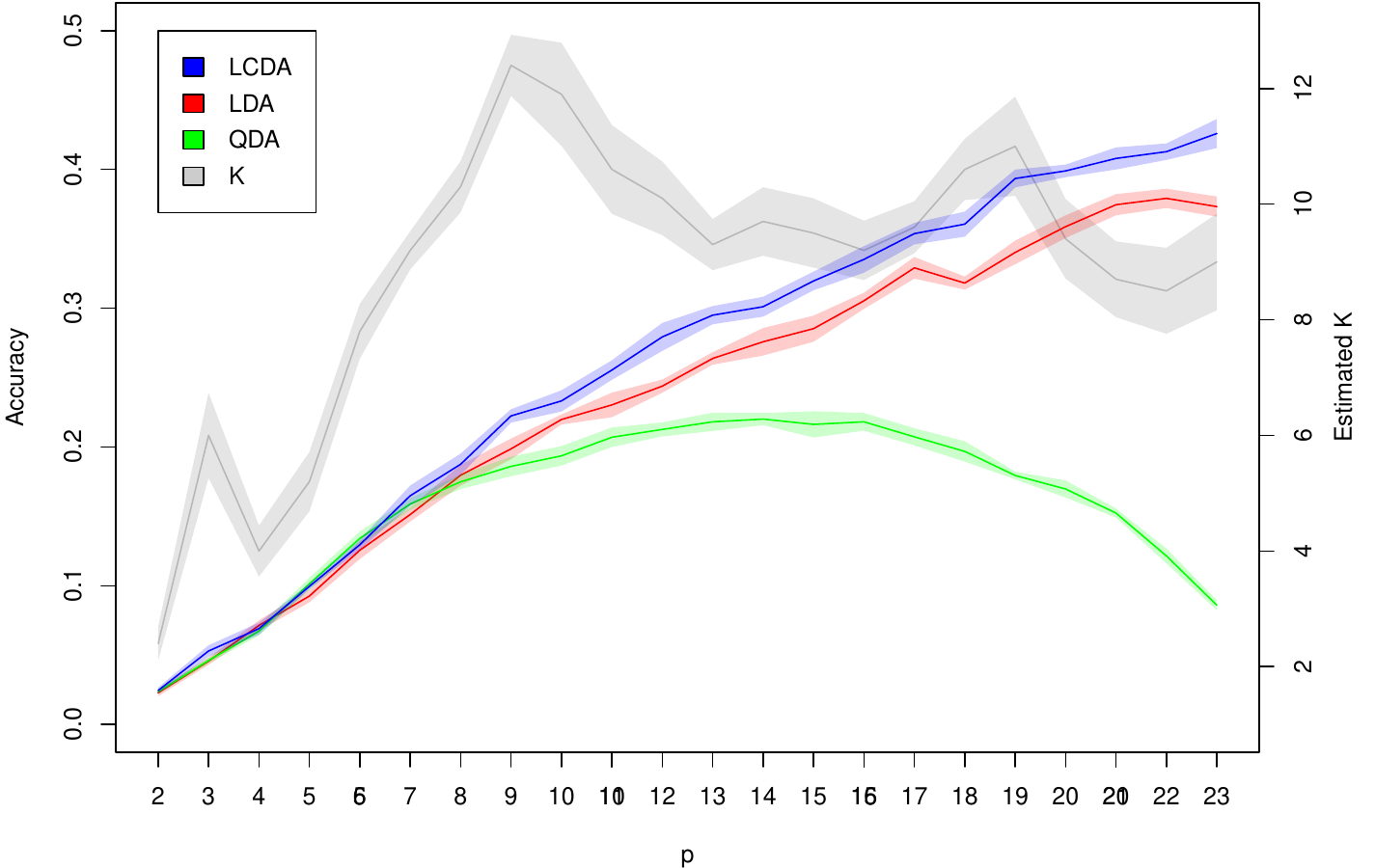} }}%
    \caption{Accuracy from the Kannada dataset comparing LCDA, LDA, and QDA.}%
    \label{fig:kannada_results}%
\end{figure}

To compare LCDA and LDA, $g$, for $g=1,2, \dots, 15$,  samples are randomly removed from each class to form the testing dataset. The models are estimated based on the training data, and class predictions are made for the left-out data. This is performed $10$ times, and the accuracy is computed. The average accuracy is calculated over the $10$ repeats and along with the $95\%$ point-wise confidence intervals. The results can be seen in \autoref{fig:kannada_results}(a). LCDA dominates LDA in classification accuracy even as the number of samples per class in the training dataset is small. It can also be seen that as the number of observations per class decreases, so does the estimate of $K$ (right y-axis of the plot). Next, taking $g=1$, the dimensions of the data from PCA are varied from $p = 2 \dots, 23$, and results are plotted in \autoref{fig:kannada_results}(b). Since the dimensions are lower than the number of samples, LCDA can be compared with both LDA and QDA. Here, we again see LCDA dominating both methods and QDA degrading as the number of dimensions increases due to unstable estimates.

% \begin{figure}%
%     \centering
%     \subfloat[\centering class 1]{{\includegraphics[width=0.12\textwidth]{kannada_examples/img001-001.png} }}%
%     \hspace{0.1cm}
%     \subfloat[\centering class 1]{{\includegraphics[width=0.12\textwidth]{kannada_examples/img001-002.png} }}%
%     \hspace{1cm}
%     \subfloat[\centering class 2]{{\includegraphics[width=0.12\textwidth]{kannada_examples/img002-001.png} }}%
%     \hspace{0.1cm}
%     \subfloat[\centering class 2]{{\includegraphics[width=0.12\textwidth]{kannada_examples/img002-002.png} }}%
%     \hspace{1cm}
%     \subfloat[\centering class 3]{{\includegraphics[width=0.12\textwidth]{kannada_examples/img003-001.png} }}%
%     \hspace{0.1cm}
%     \subfloat[\centering class 3]{{\includegraphics[width=0.12\textwidth]{kannada_examples/img003-002.png} }}%
%     \caption{First 2 images from the first 3 classes of the Kannada dataset}%
%     \label{fig:Kannada_examples}%
% \end{figure}

\section{Discussion}
\label{Discussion}

This paper proposed a novel method for clustering singular and non-singular covariance matrices using model-based clustering and extended it for classification. The situations that were targeted are a few-shot learning problems in which there are a large number of classes as well as few observations living in a high-dimensional space. The main challenge of clustering singular and non-singular covariance matrices is that they do not live in the same probability space. We first considered the case where each covariance matrix was non-singular and derived estimates via a mixture of Wishart distributions. Next, we considered the case where each matrix was singular and of the same rank and derived estimates via a mixture of singular Wishart distributions. Finally, we proposed a solution to finding parameter estimates when there is a mix of singular and non-singular matrices. This was done via mixtures of Gaussian distributions, which solved the issue of observations living in different probability spaces but at the cost of the estimated parameters being asymptotically inconsistent; a correction to this was proposed in order to have a consistent estimate. Finally, the optimal Bayes decision rule was derived for classification. The method proposed for use in classification is referred to as latent covariance discriminant analysis (LCDA) and is contrasted with LDA and QDA. Results from the simulation study show a significant improvement in terms of classification accuracy over LDA and QDA when there are latent covariance matrices in the underlying data-generating process. The proposed method was also applied to two real datasets in which the underlying data-generating process is unknown. The first was a glass dataset in which there are $200$ classes, $4$ observations per class, and a $7$ dimensional space. The second dataset is a Kannada handwritten character dataset, which consisted of $657$ classes, $25$ observations per class, and a $30$ dimensional space after feature extraction. In both of these cases, it was shown that LCDA provided a significant boost in terms of classification accuracy.

\bibliography{references1-2}

\end{document}